\PassOptionsToPackage{table}{xcolor}
\documentclass[sigconf]{acmart}
\AtBeginDocument{%
  }

\setcopyright{acmlicensed}
\copyrightyear{2026}
\acmYear{2026}
\setcopyright{cc}
\setcctype{by}
\acmConference[UIST '26]{The 39th Annual ACM Symposium on User Interface Software and Technology}{November 02--05, 2026}{Detroit, MI, USA}
\acmBooktitle{The 39th Annual ACM Symposium on User Interface Software and Technology (UIST '26), November 02--05, 2026, Detroit, MI, USA}
\acmDOI{10.1145/3830398.3830502}
\acmISBN{979-8-4007-2856-3/2026/11}

\usepackage{booktabs}
\usepackage{multirow}
\usepackage{array}
\usepackage{graphicx}
\usepackage{makecell}
\usepackage{enumitem}

\newcommand{\re}[1]{{#1}}
\makeatletter
\@ifundefined{iflatexml}{\newif\iflatexml\latexmlfalse}{}
\makeatother

\iflatexml
  \renewcommand{\authornote}[1]{}
\fi

\definecolor{brandred}{HTML}{CCA179}

\definecolor{brandred75}{HTML}{D9B99B} % 75%
\definecolor{brandred74}{HTML}{DAB99C} % 74%
\definecolor{brandred71}{HTML}{DBBC9F} % 71%
\definecolor{brandred66}{HTML}{DEC2A6} % 66%
\definecolor{brandred65}{HTML}{DFC3A8} % 65%
\definecolor{brandred63}{HTML}{DFC5AA} % 63%
\definecolor{brandred55}{HTML}{E3CCB6} % 55%
\definecolor{brandred41}{HTML}{EAD9C9} % 41%
\definecolor{brandred35}{HTML}{EDDFD1} % 35%
\definecolor{brandred32}{HTML}{EFE2D5} % 32%
\definecolor{brandred21}{HTML}{F5ECE4} % 21%
\definecolor{brandred11}{HTML}{F9F5F1} % 11%
\definecolor{brandred8}{HTML}{FAF8F5}  % 8%
\definecolor{brandred5}{HTML}{FCFAF9}  % 5%

\begin{document}

\title{TailorCoPilot: Enabling Agentic Pattern Making with Version-Controlled State Tracking}

\author{Yuexin Sun}
% \authornote{Also affiliated with College of Fashion and Design and Key Laboratory of Clothing Design \& Technology}
\affiliation{%
  \institution{Donghua University}
  \city{Shanghai}
  \country{China}}
\email{ysun@mail.dhu.edu.cn}

\author{Zhaohui Wang}
% \orcid{0000-0002-1925-7573}
\authornote{Corresponding author.}
\affiliation{%
  \institution{Donghua University}
  \city{Shanghai}
  \country{China}}
\email{wzh_sh2007@dhu.edu.cn}

\author{Ruiyang Liu}
% \orcid{0000-0003-0075-6230}
\authornote{Project leader. \\ Yuexin Sun, Zhaohui Wang are affiliated with  College of Fashion and Design, Donghua University, and Key Laboratory of Clothing Design \& Technology, Ministry of Education. This research work is conducted at \href{https://www.style3d.com/}{Style3D Research}.}
\affiliation{%
 \institution{Style3D Research}
 \city{HangZhou}
 \country{China}}
\email{novemree@gmail.com}

\settopmatter{authorsperrow=4}
\author{Demian Kong}
\affiliation{%
  \institution{Style3D Research}
  \city{Hangzhou}
  \country{China}}
\email{kongdemian@linctex.com}
  
\author{Qian He}
\affiliation{%
  \institution{Style3D Research}
  \city{HangZhou}
  \country{China}}
\email{heqianhailie@gmail.com}

\author{Gaofeng He}
\affiliation{%
  \institution{Style3D Research}
  \city{HangZhou}
  \country{China}}
\email{hegaofeng@linctex.com}

\author{Huamin Wang}
\affiliation{%
  \institution{Style3D Research}
  \city{HangZhou}
  \country{China}}
\email{wanghmin@gmail.com}

\renewcommand{\shortauthors}{Sun et al.}

\begin{abstract}

Experience-driven manufacturing, such as garment pattern making, faces a severe generational skills gap because its core expertise relies on undocumented tacit knowledge forged through day-to-day practice. 
To address this challenge, we present TailorCoPilot, an agentic pattern-making system built upon a specially designed version-control backend TailorTrace. 
TailorTrace models sewing patterns as structured, discrete states and records their transformations during the pattern-making process as explicit operation sequences defined upon the geometry primitives in the sewing pattern (panels, edges, vertices and stitches). Integrated into a conventional pattern-making GUI, TailorTrace enables seamless documentation of senior experts' tacit pattern-making knowledge without breaking their daily workflow. The documented knowledge further offers interactive, pedagogical scaffolding for novices, while providing a robust foundation to power TailorCoPilot and train future generative AI models. 
In a user study with novices and advanced novices, TailorCoPilot improved task completion rates, reduced time and perceived workload, and yielded higher-quality artifacts compared to skill-appropriate baselines. 
Ultimately, TailorCoPilot demonstrates a viable pathway to capture practice-based expertise, operationalizing it to support both generative AI advancements and human apprenticeship.
\end{abstract}



\begin{teaserfigure}
  \centering
  \includegraphics[width=\linewidth]{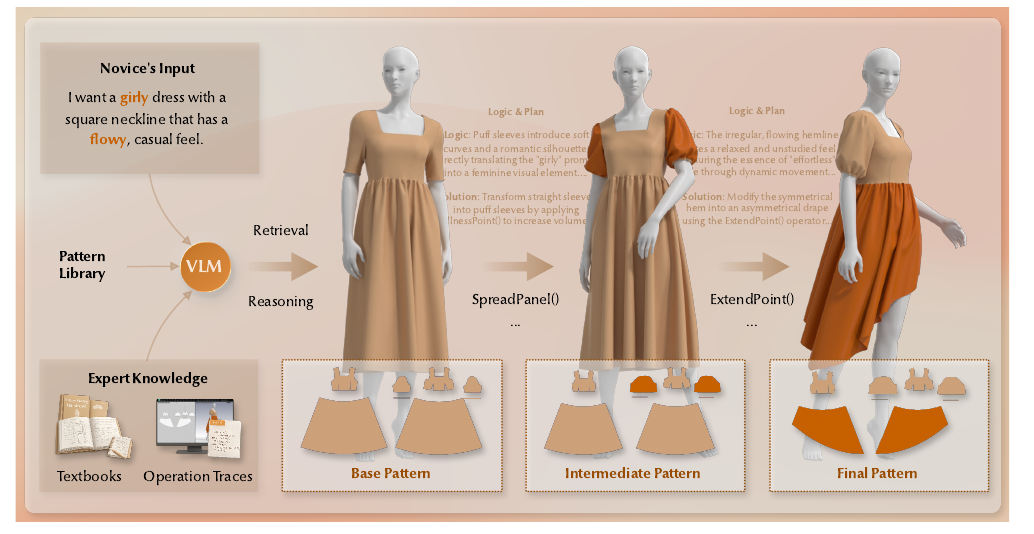}
  \caption{Overview of TailorCoPilot for novice garment pattern making. Given a novice’s natural-language design intent, TailorCoPilot retrieves a compatible base pattern from a pattern library and uses a vision-language model (VLM) to reason based on embedded expert knowledge from textbooks and operation traces. It then proposes stepwise editable pattern operations, such as \texttt{SpreadPanel()} and \texttt{ExtendPoint()}, that progressively transform the base pattern into intermediate and final patterns, with corresponding changes in the 3D garment form.}
  \label{fig:teaser}
  \Description{Overview of TailorCoPilot for novice garment pattern revision. A novice provides a natural-language design request, the system retrieves a compatible base pattern, reasons over expert knowledge from textbooks and operation traces with a vision-language model, and applies editable operations to transform the base pattern into intermediate and final patterns with matching 3D garment updates.}
\end{teaserfigure}

\maketitle

\section{Introduction}

Experience-driven manufacturing domains rely heavily on expertise built through repeated practice and only rarely formalized in textbooks~\cite{Seghroucheni2023SystematicRO}. 
As senior practitioners retire and fewer newcomers enter these trades, a large body of practical knowledge faces the risk of being lost before it can be passed on~\cite{Guo2025, Kaipainen2025, Partarakis2025}.
This problem is especially visible in garment pattern making. Experienced pattern makers develop judgment through years of hands-on work, learning how to translate design intent into production-ready pattern structures while preserving fit, balance, and sewability~\cite{Hameed2020, Lyu2025}. \re{Much of this knowledge is embedded in procedural habits, intermediate decisions, and localized repair strategies, making it nearly impossible to reverse-engineer from finished artifacts alone~\cite{Jo2022GenerativePP}. Addressing this gap requires a dedicated, trace-capturing infrastructure to document the nuanced steps of expert workflows---an objective that serves as the core motivation for this research.}

\re{Existing computational tools mostly focus on modeling final geometric outcomes, while failing to capture the crucial intermediate steps of pattern making. Professional Computer-Aided Design (CAD) systems provide robust precision and editing power, but they function largely as blank canvases. They operate under the assumption that users have already mastered the tacit drafting logic and established conventions necessary to build a pattern from scratch~\cite{Xiu2013}. To lower this learning barrier, some providers have introduced parametric, template-driven systems; however, these inherently sacrifice creative degrees of freedom by restricting designers to predefined structural boundaries. More recently, generative AI has attempted to automate the craft by predicting system parameters or directly outputting final vector layouts~\cite{Gui2025GenPatternDE, Luo2025DeepLF}. Yet, due to the probabilistic nature of generative models, their outputs struggle to consistently meet the strict geometric tolerances required for industrial manufacturing, while the absence of intermediate drafting states means that human practitioners lack the procedural context needed to interpret, adjust, or locally repair the AI-generated results.}

To address these challenges, we introduce \textbf{TailorTrace}, a version-controlled computational framework designed to track and structure state changes throughout garment pattern development. By representing sewing patterns as discrete states and recording transitions as explicit geometric operations, TailorTrace renders the inherently opaque pattern-making process legible at the level of states, actions, and revisions. This architecture enables expert workflows to be captured as interpretable traces, revisited step-by-step, and reused as operational resources. Crucially, by documenting tacit expert knowledge as explicit operational traces, TailorTrace establishes a robust data foundation for future generative AI, allowing models to learn from expert-defined transformations rather than relying solely on static, final-outcome examples.

Building upon this trace-capturing infrastructure, we present \textbf{TailorCoPilot}, an agentic co-creation system that leverages these recorded workflows to provide state-aware, interactive assistance. During the drafting process, TailorCoPilot actively surfaces relevant revision histories and intermediate structures, offering essential procedural context and guided scaffolding for digital apprenticeship~\cite{Valdemar2023}. We evaluate TailorCoPilot through a user study involving novice and advanced-novice pattern makers, comparing its efficacy against baseline tools matched to their respective skill levels. Our results demonstrate quantitative and qualitative improvements in task completion rates, time efficiency, perceived workload, and final artifact quality. We summarize the primary contributions of this paper as follows:

\begin{itemize}
    \item A process-centered framing of garment pattern making as traceable iterative revision.
    \item TailorTrace, a CAD-integrated, version-controlled representation of validated pattern states and domain-specific transformations\footnote{\re{The current evidence is limited to bounded pattern-revision tasks in a controlled setting}}.
    \item TailorCoPilot, an interactive agentic system that provides stepwise, editable, and human-controllable support through traceable states.
    \item Evidence from user and diagnostic studies that trace-based support improves the performance of novices and advanced novices.
\end{itemize}

\section{Related Work}

\subsection{Garment Pattern-Making Tools}

Digital support for garment pattern making has developed along several directions. Industrial CAD systems such as Lectra and ET~\cite{Lectra25,ET25} support precise drafting and production workflows, but they typically assume substantial knowledge of geometric operations, drafting logic, and professional conventions~\cite{Xiu2013,Rahman2015,Boateng2025,Lee24}. Parametric methods further improve efficiency through measurement-driven rules and constrained variables~\cite{Liu2019ParametricDO,Jin2023DesignAR,Kim25}, and recent surveys highlight the growing role of pattern-centered workflows in digital fashion design~\cite{Lyu2025,Luo2025DeepLF}.

Recent work has explored structured representations and learning-based approaches for garment generation. GarmentCode represents sewing patterns as executable programs~\cite{Korosteleva2023GarmentCodePP}, and GarmentCodeData provides large-scale paired datasets~\cite{Korosteleva24}. SewFormer and other generative models support garment and sewing-pattern reconstruction or generation from visual, textual, or multimodal input~\cite{liu2023sewformer,Chen24,Liu24,He24,Nakayama24,Li25,garmagenet2025}. \re{These end-to-end approaches can produce complete pattern outputs, but their intermediate decisions and revision paths are often difficult to inspect, edit, or reuse.} ChatGarment~\cite{Bian24} and Design2GarmentCode~\cite{Zhou24} use VLMs to infer higher-level design parameters for constrained pattern generation. \re{Such parametric AI methods improve controllability for predefined garment variations, yet their flexibility is often bounded by the available templates and control variables. In contrast, TailorTrace and TailorCoPilot represent pattern making as traceable meta-operations over validated pattern states, supporting finer-grained revision while preserving interpretable transitions between states.}

\subsection{Capturing Expert Processes in Design Work}

A central challenge in the garment industry is that much of the relevant expertise is tacit and situated in practice. Prior work~\cite{Fatma2022,Hameed2020,Papahristou24} notes that expert knowledge often resides in judgment, collaboration, and context-sensitive decision making that is difficult to document in formal rules alone. This challenge is especially acute in pattern making, where local edits often need to be evaluated in relation to downstream construction constraints and interactions among multiple pieces.

In computational design, structured representations have been used to encode geometry, constraints, and design procedures. DeepCAD models CAD construction sequences~\cite{Wu2021DeepCADAD}, SketchGraphs~\cite{Seff2020SketchGraphsAL} represents relational sketch structure, and Vitruvion~\cite{Seff2021VitruvionAG} generates parametric CAD sketches. More recent approaches~\cite{Jones2025ASH,Casey2025AligningCG} such as solver-aided CAD languages, scene programs~\cite{Zhang2024TheSL}, and workflow-capture systems like ClearFairy~\cite{Son2025ClearFairyCC} further demonstrate that procedural steps and design rationales can be represented explicitly. However, these approaches primarily encode geometric structure, executable programs, or generic workflows, and do not address how expert revision knowledge can be \re{captured as validated, operation-level traces} for coordinated garment-pattern modification.

\subsection{Human-AI Support}

Recent HCI research increasingly frames AI systems as interactive partners rather than purely generative tools. Work on co-creative systems and human-AI collaboration highlights the importance of user agency, control, and iterative interaction~\cite{Rezwana2022COFI,Zhang25Exploring,Xi25}. Related studies~\cite{Peuter21,Wiberg23,Ding24} further emphasize transparency, task-dependent interfaces, and support for user steering during generation. More recently, agentic AI design systems have begun to incorporate multi-step reasoning, planning, and tool use in creative workflows~\cite{Acharya2025AgenticAA,Son2025ClearFairyCC}.

Recent work also focuses on making intermediate decisions inspectable and revisable. ArtKrit~\cite{Ma2025ComputationalSO} introduces computational scaffolding for structured artistic workflows, while DesignTrace~\cite{Peng2026DesignTrace}, GenTune~\cite{Chung2025GenTune}, and ImaginationVellum~\cite{Marquardt2025ImaginationVellumGI} support traceable refinement and iterative design.  \re{However, these systems target general creative refinement and do not address manufacturing-oriented revision, where each operation must preserve geometric, topological, and physical validity.} Reviews of interactive and user-centered AI argue for moving beyond post-hoc explanations toward interaction-time support~\cite{interactiveai2024,Hong2025DevelopingUS}, and work in fashion design shows that graphical interfaces can better support exploration than text prompting alone~\cite{fashioning2024}. Together, these studies highlight the importance of inspectability, controllability, and iterative refinement in human-AI design systems. \re{We extend this perspective to garment pattern making through validated operation traces that support interactive revision and novice learning.}

\section{Formative Study}

Garment pattern making is fundamentally a complex process of inverse spatial deconstruction. It converts a conceptual design, typically a 2D stylized illustration of the garment's desired 3D draped form on a human avatar, into a precise set of flat 2D geometric primitives. Unlike traditional engineering CAD where 3D modeling is largely "what-you-see-is-what-you-get", pattern making is highly indirect and requires profound spatial intuition from the pattern makers to mentally deconstruct the 3D illustration into flat pattern pieces with notch notations for assembly, while accounting for physical properties of fabrics and strict dimensional precision. For novice practitioners, cultivating this dynamic mental model purely from static textbooks is extremely difficult, not to mention that such complex 2D-to-3D geometric reasoning remains a significant blind spot for current generative AI systems.

TailorCoPilot is designed to bridge the expertise gap between novice and expert pattern makers and provide the missing infrastructure required to integrate generative AI into the complex pattern-making workflows. Before delving into TailorCoPilot, we conducted a formative study to identify how expert pattern makers manage interconnected geometric constraints between 2D patterns and 3D draping, where novices fail to anticipate downstream consequences, and what foundational barriers underlie this gap.

\subsection{Participants and Procedure}

We recruited 13 participants through purposive sampling, following the early stages of the Dreyfus model of skill acquisition~\cite{Dreyfus04}. The sample included \textbf{6 novices} (PF1--PF6) with limited or no pattern-making experience, \textbf{4 junior pattern-making students as advanced novices} (PF7--PF10) who had acquired basic pattern-making skills via textbooks, and \textbf{3 experts} (PF11--PF13) who were senior pattern makers or pattern-making educators with more than 5 years of industry experience.
Each session lasted about 60 minutes, including a 30-minute semi-structured interview and a 30-minute task session. In the interview, we asked about participants' workflows, how they learned or taught pattern-making knowledge, how they decomposed garment changes, common breakdowns, and the criteria they used to judge whether a modification was structurally and aesthetically sound.

\begin{table}[!t]
  \centering
  \caption{Representative tasks categorized by difficulty. \textbf{Simple} tasks involve local parametric refinements to base patterns, such as length or width adjustments. \textbf{Medium} tasks encompass modular component synthesis and constrained topological operations, such as dart transfers, pleating and gathering. \textbf{Complex} tasks require reference-driven deconstruction from design sketches or advanced structural transformations.}
  \label{tab:tasks_summary}
  \small
  \setlength{\tabcolsep}{8pt}
  \renewcommand{\arraystretch}{1.12}
  \begin{tabular}{>{\centering\arraybackslash}m{0.8cm} >{\centering\arraybackslash}m{0.5cm} >{\centering\arraybackslash}m{5.5cm}}
    \toprule
    \textbf{Difficulty} & \textbf{Count} & \textbf{Typical Case} \\
    \midrule

    \textbf{Simple}
    & 15
    & \begin{minipage}[t]{\linewidth}
        \centering
        \includegraphics[height=2.1cm]{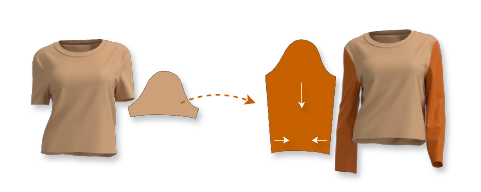}\\[-1pt]
        {\footnotesize  Convert short sleeves to long sleeves}
      \end{minipage}
    \\[5pt]

    \textbf{Medium}
    & 10
    & \begin{minipage}[t]{\linewidth}
        \centering
        \includegraphics[height=2.1cm]{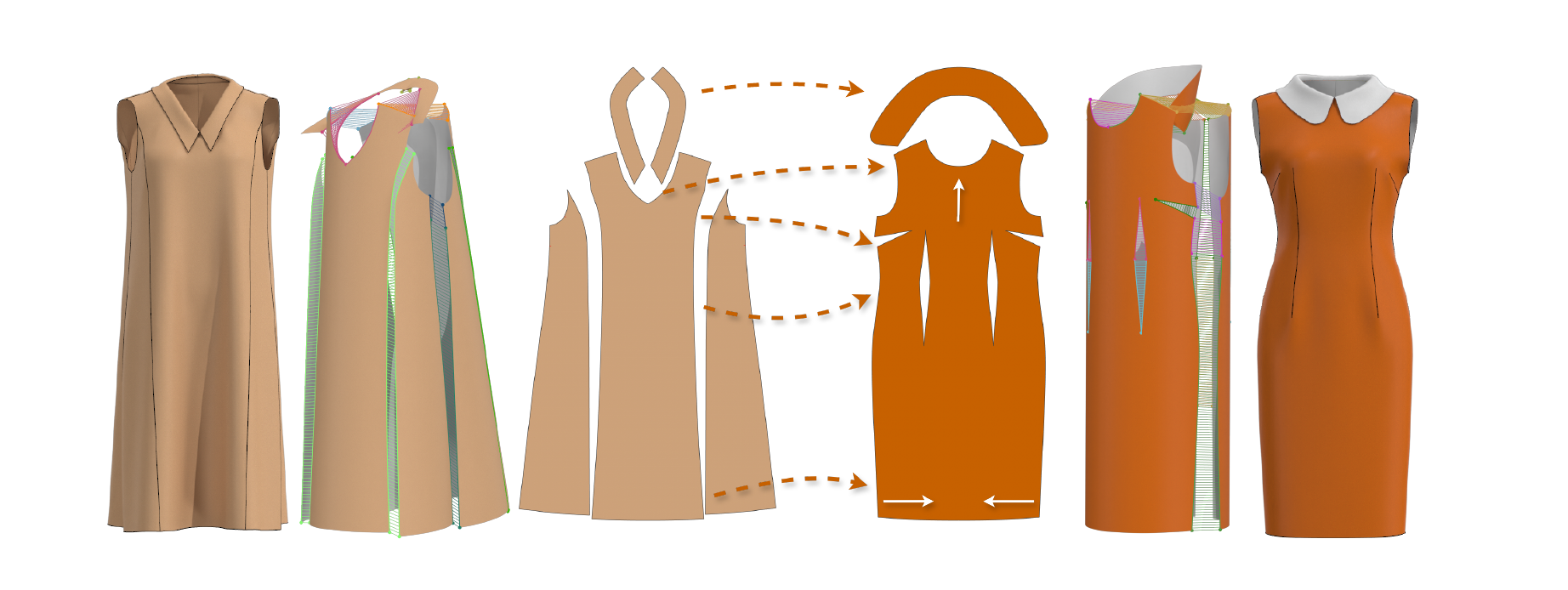}\\[-1pt]
        {\footnotesize  Change collar type, seam lines, and hems}
      \end{minipage}
    \\[5pt]

    \textbf{Complex}
    & 5
    & \begin{minipage}[t]{\linewidth}
        \centering
        \includegraphics[height=2.1cm]{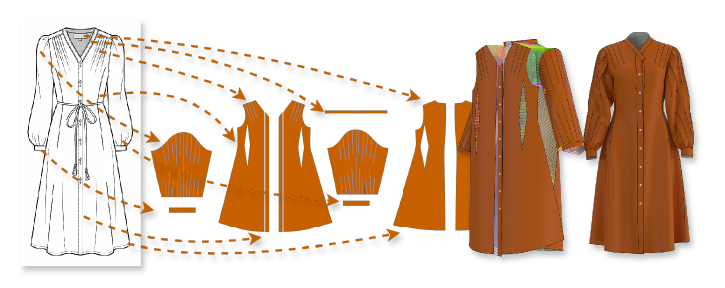}\\[-1pt]
        {\footnotesize  Create a dress from a reference image}
      \end{minipage}
    \\

    \bottomrule
  \end{tabular}
  \Description{Table of representative tasks by difficulty. Simple tasks include 15 local refinements such as converting short sleeves to long sleeves. Medium tasks include 10 component and topology changes such as changing collar type, seam lines, and hems. Complex tasks include 5 reference-driven tasks such as creating a dress from a reference image.}
\end{table}

\begin{figure*}[t]
    \centering
    \includegraphics[width=0.9\linewidth]{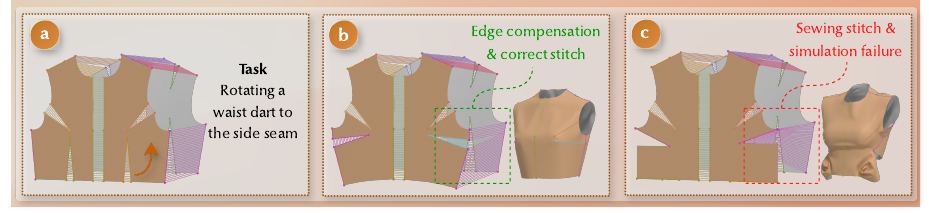}
    \caption{Example from the formative study. (a) Base patterns for a dart-rotation task. (b) Correct result, which requires both dart relocation and corresponding seam adjustment and edge compensation. (c) A novice participant's actual result. Only the dart position was changed, while the stitch relationship was left unresolved, leading to sewing and simulation failure.}
    \label{fig:formative}
    \Description{Three-panel example from the formative study showing that a correct garment revision must preserve relationships across pattern pieces. The correct result moves a waist dart to the side seam and also adjusts seam length and stitching, while the novice result changes only the dart position and causes sewing and simulation failure; a: Base bodice pattern for a dart-rotation task. An arrow indicates moving the waist dart from its original location toward the side seam; b: Correct result after dart rotation. The dart is relocated, the side seam and edge lengths are compensated, and the 3D preview shows a valid sewn bodice; c: Novice result after changing only the dart position. The stitch relationship is left unresolved, so the 3D preview shows sewing and simulation failure.}
\end{figure*}

In the task sessions, participants performed representative tasks across three difficulty levels (Simple, Medium, and Difficult), encompassing operations ranging from local refinements to structural modifications and reference-driven creations (Table~\ref{tab:tasks_summary}). To ensure ecological validity, these tasks were validated by expert educators to mirror authentic practitioner workflows, specifically mapping design initiation, progression trajectories, and the exact points where participants encountered friction. During the session, participants were allowed to use their preferred CAD environments, supplemented by Style3D~\cite{Style3D} for virtual draping and AI-based pattern assistants~\cite{Nakayama24, Zhou24}. To minimize cognitive interference during the session, we employed retrospective think-aloud protocols~\cite{Haak2003RetrospectiveVC} and screen recordings to isolate critical decision points, workflow breakdowns, and repair strategies.

We collected interview transcripts, task recordings, and resulting 2D and 3D artifacts, and analyzed them using reflexive thematic analysis~\cite{braun06}. Preliminary clustering was used to organize recurring issues in the textual data~\cite{xu2025tama}, after which two researchers reviewed the materials, refined the themes, and derived implications for system design.

\subsection{Findings}

Our formative study yielded four findings about how pattern making begins, why expert knowledge remains hard to access, where less experienced users struggle, and how participants judge whether a revision is working. Together, these findings motivate a system that starts from an initial base pattern, digitizes expert pattern-making knowledge into reusable traces, and keeps revision inspectable and revisable.

\subsubsection{Workflow}
\label{finding:workflow}
Among all difficulty levels, unless necessary, participants rarely draft sewing patterns from scratch. Instead, they usually started from a pattern close to the target and modified it through comparison. Advanced novices and experts often drew this starting point from a pattern library or AI-assisted generation, then continued editing in CAD. Novices lacked such resources and could not operate CAD, so they relied almost entirely on end-to-end AI generation. As one expert explained, ``We usually begin from the closest block we can find, then modify toward the target'' (PF11). This suggests that revision support should first establish an editable base pattern.

\subsubsection{Digitizing Expert Knowledge}
\label{finding:knowledge}
Experts did not frame pattern making as fixed rules alone. They combined structural principles, prior cases, embodied judgment, and process habits to decide what to change, in what order, and whether a result remained sound. For less experienced users, this knowledge was hard to access because it was scattered across textbooks, teaching, examples, and personal practice. Revisions were also path-dependent, since later decisions relied on earlier ones and experts often reasoned through intermediate states. As one participant noted, ``The know-how is not just the rule itself, but knowing when to use it, what to adjust next, and how to tell if it is going wrong'' (PF13). This suggests that a central design challenge is how to digitize expert knowledge as reusable traces, decision criteria, and revision records, rather than preserving only final outcomes.

\subsubsection{Barriers From Intents to Edits}
\label{finding:barriers}
A central difficulty was turning intended appearance into executable edits. Novices could often describe a desired effect, but could not determine what part of the pattern to change or what operation would produce it. For example, one participant wanted a silhouette that felt ``softer and more expanded'' (PF2), but could not identify the geometric changes needed for the sewing pattern. Advanced novices could understand concepts such as dart transfer and perform local edits to the sewing pattern, yet still struggled with precision, sequencing, and more complex modifications. As one participant put it, ``I can describe the style I want, but I still do not know what to adjust first on the pattern'' (PF5). These observations indicate that expert support becomes useful only when intent can be translated into inspectable steps and editable parameters.

\subsubsection{3D Feedback and Intervention}

Novice participants generally relied heavily on 3D try-on and other visual feedback to judge whether a revision was working. They often could not assess the consequences of a local 2D change until sewing or simulation exposed them. For example, one novice (PF3) completed a dart change without resolving the corresponding seam relationship, leading to simulation failure (Figure~\ref{fig:formative}). Experts used intermediate visualizations to check proportion, balance, seam compatibility, and wearing effect, while less experienced users needed to inspect the evolving result before deciding what to do next. As one advanced novice remarked, ``I need to see the garment update step by step before I can trust the modification'' (PF8). This suggests that support tools should allow users to inspect and intervene in the revision process itself.

\begin{figure*}[t]
    \centering
    \includegraphics[width=\textwidth]{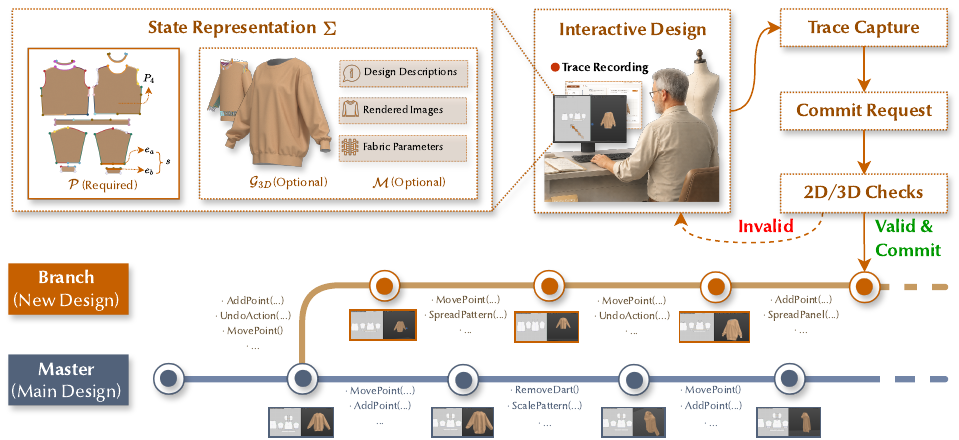}
    \caption{\re{Overview of TailorTrace. Experts begin by exploring and modifying garment designs through an interactive frontend. As they edit, the backend captures their semantic operations as traces and maintains each design state using a unified representation that includes required 2D pattern geometry, optional 3D draping results, and multimodal metadata. Once a design reaches a milestone stage, the expert can request a commit. TailorTrace first performs checks on the 2D pattern and, when 3D geometry is available, further evaluates the draping result. A valid state is committed as a new revision in a branching history, allowing experts to revisit earlier states and continue exploring alternative design directions. An invalid state is returned to the frontend for further editing before another commit is attempted.}}

    \label{fig:support}
    \Description{Overview of the TailorTrace workflow for version-controlled garment design. The workflow begins with an expert interactively editing a garment pattern in the frontend. The backend records the resulting semantic editing operations as traces and stores each design using a unified state representation. Every state contains required 2D pattern geometry and may additionally include 3D draping geometry, design descriptions, rendered images, and fabric parameters. When the expert submits a commit request, TailorTrace checks the 2D pattern and, if 3D geometry is available, also checks the draping result. States that pass validation are added to a branching revision history, where experts can revisit previous revisions and explore alternative branches. States that do not pass validation are returned to the frontend for continued editing and later resubmission.}

\end{figure*}

\section{System Design}
\label{sec:system}

Guided by our formative study, we introduce TailorTrace (Sec.~\ref{sec:tailortrace}), a novel interaction paradigm for pattern making that bridges tacit, physical-world pattern-making expertise with a structured, machine-readable version control backend. Designed as a transparent layer beneath a traditional CAD interface, TailorTrace captures discrete semantic actions, rigorously validates physical constraints, and records these traces as multi-modal commits. Apart from facilitating knowledge transfer between senior and junior pattern makers, the high-quality, structured trajectory data recorded with TailorTrace further enables the design of our agentic pattern-making system TailorCoPilot (Sec.~\ref{sec:tailorcopilot}). \re{Further details of the system design are provided in Supplementary Section~2.}

\subsection{TailorTrace Design}
\label{sec:tailortrace}

\subsubsection{Unified State Representation.} Unlike traditional software version control where intermediate commits may contain broken or non-compiling code, TailorTrace enforces a strict ``Valid State'' paradigm. Every commit represents a functionally viable garment-pattern state modeled as a structured discrete state encompassing both geometric primitives and rich semantic metadata\re{, with draping information included when available}. Formally, we define a garment state as a tuple $\Sigma = \langle \mathcal{P}, \mathcal{G}, \mathcal{M} \rangle$ involving:
\begin{itemize}[leftmargin=*]
    \item \textbf{Structured 2D Geometry} ($\mathcal{P}=(P,S)$) with a set of 2D panels $P = \{p_1, \dots, p_N\}$ and stitches $S$. Each panel $p_i = (e_{i,1}, \dots, e_{i,k})$ is a closed loop of curved edges while each stitch $s \in S$ is defined upon a pair of compatible edges $(e_a, e_b, \rho_s)$ with parameters $\rho_s$.
    \item \textbf{3D Draped Geometry} ($\mathcal{G}$) of the pattern, conditionally instantiated (i.e., $\mathcal{G} \neq \emptyset$) if and only if the initial state or the user's subsequent workflow incorporates 3D simulation.
    \item \textbf{Multi-Modal Metadata} ($\mathcal{M}$): A rich auxiliary representation that encapsulates design descriptions of the garment, rendered images of the 3D garment and physical fabric parameters (e.g., density, bending, and shearing stiffness) if available.
\end{itemize}

\subsubsection{Semantic Action Space and Transformations.} To digitize tacit pattern-making knowledge, TailorTrace follows neuro-symbolic design and tracks semantic design intent through explicit transformation sequences~\cite{Bhuyan2024}.
We define a comprehensive action space over the geometric primitives, such as \texttt{AddPoint(...)}, \texttt{SpreadPanel(...)}, or \texttt{MovePoint(...)} with parameters defined by mouse or keyboard input events. By recording the transformation between commits, the system captures the ``how'' and ``where'' of a pattern's evolution, creating a highly interpretable revision history.

\subsubsection{Validation Engine.} 
% To maintain the integrity of the "Valid State" paradigm, we implemented a robust, multi-tiered evaluator that operates as a continuous pre-commit hook within TailorTrace. Before any transformation sequence is committed to the version history, it must pass a series of deterministic physical and topological checks. In 2D sewing patterns, the evaluator verifies that the proposed sewing pattern maintains closed loops, contains zero intersecting geometry, and exhibits valid topological stitching (e.g., no repeated stitches or physically infeasible seams). When 3D simulation is enabled, the evaluator performs a preliminary draping simulation onto a target avatar, checking for severe self-intersections, inverted normals, or obvious body collision. Commits that violate these constraints are flagged and passed to the user for correction.
\re{To maintain the integrity of the ``Valid State'' paradigm, we implemented a robust, multi-tiered evaluator that operates at commit time rather than after each individual edit. Users may freely explore and repair intermediate configurations between commits. TailorTrace stores a state only after it passes validation, together with the full sequence of semantic operations that produced it, including repair operations. This design supports reliable revision histories and curated training trajectories for downstream learning, while excluding abandoned or unresolved invalid configurations as standalone states.

In the 2D domain, the evaluator verifies that all contour loops are closed, no geometric self-intersections are present, and topological stitching is valid (e.g., no repeated stitch assignments or unmatched seam edges). When 3D simulation is enabled, the evaluator performs a preliminary draping simulation onto a target avatar, checking for severe self-intersections, inverted normals, or obvious body collision. The 3D checks use the simulator's default collision and normal-validity tolerances. Commits that violate these constraints are rejected, and users can continue editing until the state becomes valid.}

\subsubsection{TailorTrace Interaction Design.} A core interaction goal of TailorTrace is \re{to introduce minimal additional friction into existing workflows}~\cite{Sweller1988CognitiveLD}. Senior pattern makers possess deep tacit knowledge but often resist disruptive shifts in their tooling. Therefore, the version control mechanics (branching, committing, rebasing) are abstracted away and embedded invisibly behind a standard CAD GUI.
Users maintain their existing drafting workflows, utilizing familiar input modalities like standard mouse peripherals or stylus interfaces, while TailorTrace records these workflows as traces in the background, ensuring that cognitive load remains entirely on the creative task rather than on managing version control trees.

\begin{figure*}[!t]
    \centering
    \includegraphics[width=\linewidth]{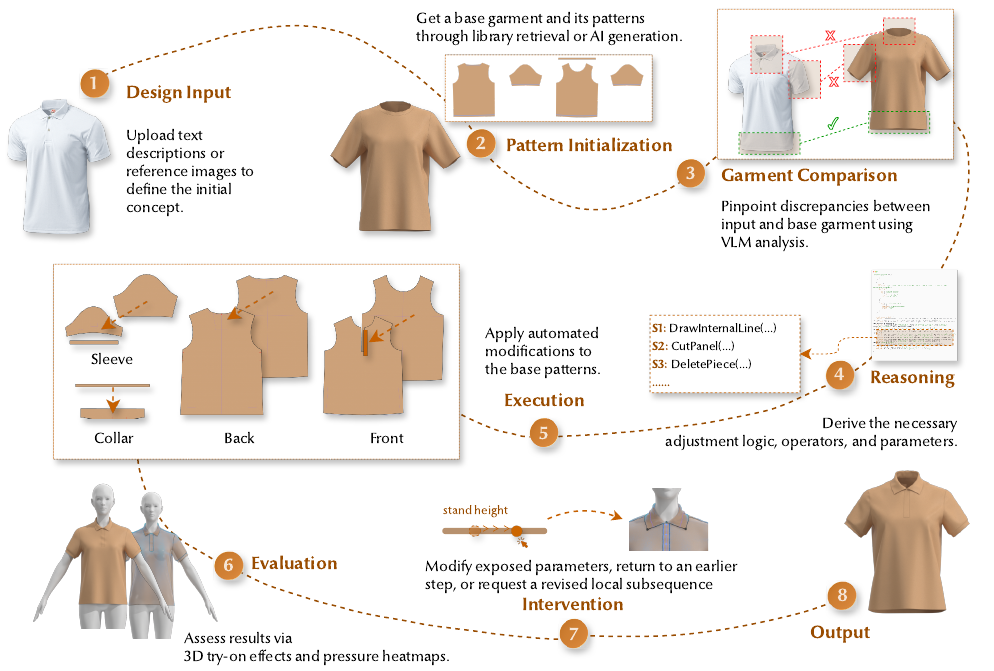}
    \caption{Novice walkthrough of TailorCoPilot. The workflow proceeds through eight stages: (1) \textbf{Design Input}, where the user provides a text description or reference image; (2) \textbf{Pattern Initialization}, where the system gets a base garment and its patterns by retrieving from a user-provided pattern library or AI generation; (3) \textbf{Garment Comparison}, where a VLM pinpoints discrepancies between the input and the retrieved garment; (4) \textbf{Reasoning}, where the system derives adjustment logic, symbolic operators, and editable parameters; (5) \textbf{Execution}, where the proposed modifications are applied to the base pattern; (6) \textbf{Evaluation}, where results are assessed through synchronized 3D try-on effects and pressure heatmaps; (7) \textbf{Intervention}, where users can modify exposed parameters, return to earlier steps, or request a revised local subsequence; and (8) \textbf{Output}, the final revised garment pattern and corresponding garment form.}
    \label{fig:walkthrough}
    \Description{Eight-stage novice workflow of TailorCoPilot. The user provides text or reference images, the system initializes a base pattern, compares it with the target garment, reasons about needed edits, executes symbolic modifications, evaluates the result with 3D try-on and pressure heatmaps, allows user intervention, and outputs the final revised garment pattern and garment form.}
\end{figure*}

\begin{figure*}[!t]
    \centering
    \includegraphics[width=0.9\linewidth]{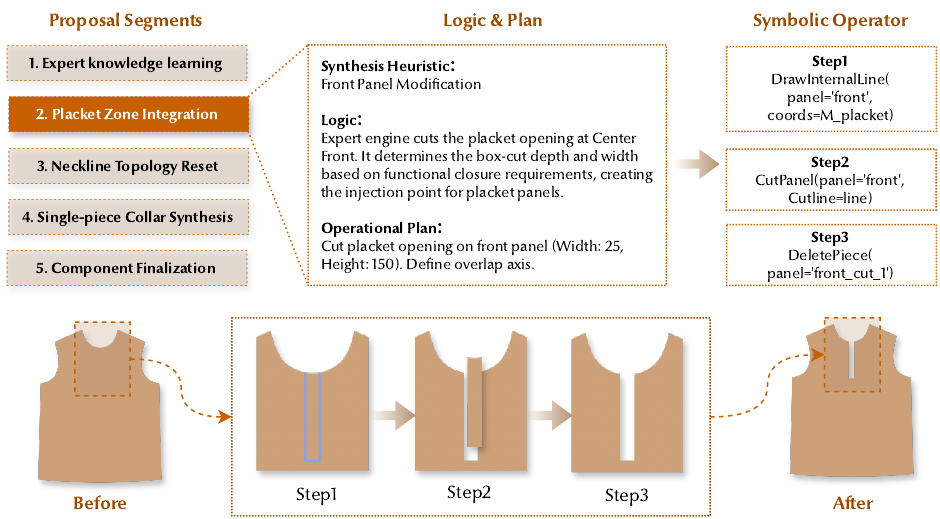}
    \caption{Neuro-symbolic reasoning and execution flow. The system translates design intent into structural heuristics, compiles them into an operational plan, and executes symbolic edits on the pattern topology. In the illustrated placket integration step, the engine determines the center-front opening logic before applying the corresponding backend operators.}
    \label{fig:reasoning}
    \Description{Neuro-symbolic reasoning and execution flow for integrating a placket into a front panel. The system converts design intent into structural logic, compiles an operational plan, and applies three symbolic editing operators to create the center-front opening, with before-and-after panels showing the resulting topology change.}
\end{figure*}

\subsection{TailorCoPilot: Agentic Pattern Making}
\label{sec:tailorcopilot}

Leveraging the structured semantic traces captured by TailorTrace, we developed TailorCoPilot, an agentic pattern-making system capable of translating unstructured design intent into executable operations. To train this system, we curated a dataset of 93 high-quality design traces, comprising 56 foundational traces sourced from pattern-making textbooks (spanning 3 to 10 states each) and 37 real-production traces \re{captured live through TailorTrace as industry experts worked in standard CAD software, rather than reconstructed afterward} (spanning 10 to 30 states each). \re{TailorTrace ran as a background CAD plugin that automatically recorded editing operations, while explicit trace-management actions such as commits and branches were performed through button clicks or keyboard shortcuts.}

To prevent data leakage and \re{evaluate operation synthesis on held-out traces}, we strictly maintained a trace-level split for our training and validation sets. Within the training traces, we employed a distance-based stratified sampling strategy to capture a diverse distribution of both short-range, atomic edits and long-range, compound modifications. For each sampled state pair, we utilized Gemini~\cite{gemini,comanici2025gemini} to generate unstructured, natural language descriptions of the visual and intentional design differences\re{, which domain experts manually verified and corrected when necessary for technical accuracy and authentic terminology}. By pairing these natural language descriptions with the exact, discrete operation sequences extracted from TailorTrace, we constructed a comprehensive dataset of approximately 2,500 translation pairs. Finally, we utilized this dataset to fine-tune the Qwen3-VL-8B~\cite{qwen3technicalreport} model, yielding the core reasoning engine for TailorCoPilot.

Figure~\ref{fig:walkthrough} shows the overall workflow of TailorCoPilot. A user begins by presenting their design intent through either a textual description of the garment style or a reference image/sketch. In response to Finding~\ref{finding:workflow}, TailorCoPilot first initializes a base pattern either via generative model or by retrieving one from a provided pattern library with VLM~\cite{qwen3vlembedding} extracted features. The system then compares the initialized pattern against the user’s intent and identifies the pattern components that require change.

Given the current pattern state and relevant expert resources, the reasoning module then analyzes the desired modifications to be made and proposes a sequence of stepwise editable operations, where each step specifies what part of the pattern should change, which operator should be applied, and what parameters should be adjusted in the current context, as shown in Figure~\ref{fig:reasoning}. These operations progressively transform the base pattern into intermediate and final states, with corresponding changes shown in the 3D garment form. Throughout this process, the interface exposes the current logic, editable parameters, and resulting 2D pattern changes so that novice users can inspect, replay, and intervene in the revision process to enhance their own knowledge. 
\begin{figure}[!t]
    \centering
    \includegraphics[width=0.92\linewidth]{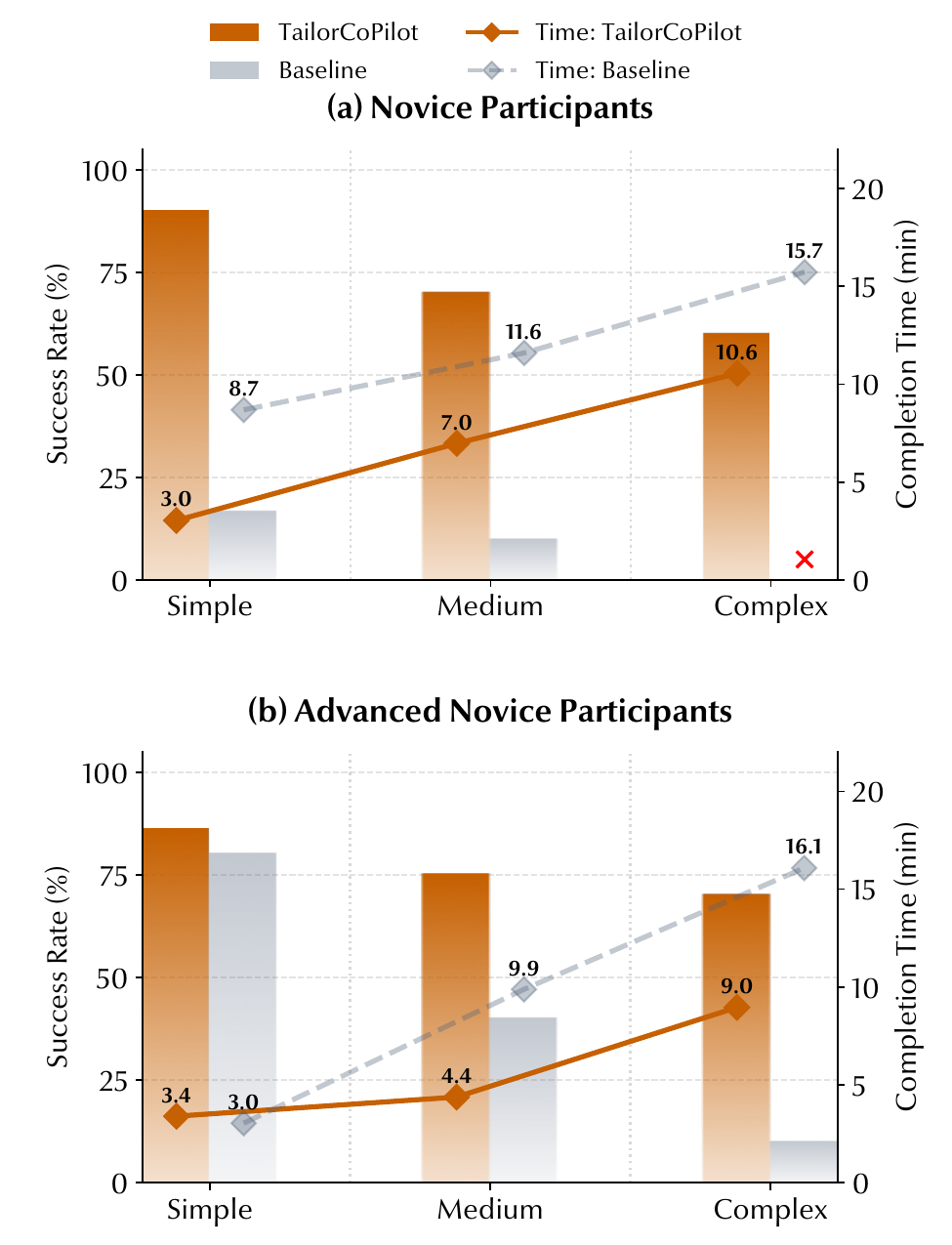}
    \caption{Task success and completion time by workflow, difficulty, and expertise group. For novices, TailorCoPilot maintained a clear success-rate advantage across all task difficulties. For advanced novices, TailorCoPilot and the baseline were closer on simple tasks, while TailorCoPilot achieved higher success and lower completion time on medium and complex tasks.}
    \label{fig:eval_performance}
    \Description{Comparison of TailorCoPilot and a baseline on task success and completion time across simple, medium, and complex tasks. TailorCoPilot improves success for novices at all difficulty levels and gives the clearest gains on medium and complex tasks for advanced novices, while also reducing completion time on harder tasks; a: results for novice participants. TailorCoPilot has substantially higher success rates than the baseline for simple, medium, and complex tasks, and its completion times remain lower, especially on complex tasks; b: results for advanced novice participants. TailorCoPilot and the baseline are close on simple tasks, but TailorCoPilot achieves higher success and lower completion time on medium and complex tasks.}
\end{figure}

\section{Evaluation}

We evaluate our system from two complementary perspectives to validate both the interactive benefits of TailorCoPilot and the foundational utility of TailorTrace. First, we conduct a user study to assess whether the trace-based support embedded within the TailorCoPilot system improves interactive pattern making for novice users, evaluating its capacity to close the generational knowledge gap between experts and beginners. \re{We then conduct a diagnostic study comparing VLM-only, textbook-trace-only, and expert-trace-augmented models, demonstrating that TailorTrace captures critical expert knowledge that lies beyond static LLM knowledge and textbook materials.}

\subsection{Evaluation Settings}

To avoid carry-over effects from the formative study, we recruited a new pool of 20 participants, including \textbf{10 novices} (PU1--PU10) and \textbf{10 advanced novices} (PU11--PU20). We used a mixed design with \emph{Workflow} (TailorCoPilot vs.\ baseline) and \emph{Difficulty} (simple, medium, complex) as within-participant factors, and \emph{Expertise} (novice vs.\ advanced novice) as a between-participant factor. As in the formative study, participants were allowed to choose the baseline pattern-making tools they could use, including AI~\cite{Nakayama24, Zhou24} and CAD~\cite{Style3D}.

Each participant completed 12 trials across two sessions held on separate days, with at least one day between sessions. Each session included 6 trials, with 3 simple, 2 medium, and 1 complex task. Trials were divided into two 3-trial blocks separated by a short break. Workflow order was counterbalanced across participants using an AB/BA scheme at the block level to reduce context switching between workflows. Each participant completed 6 simple, 4 medium, and 2 complex trials in total. Within each difficulty level, trials were split evenly between TailorCoPilot and the baseline. Task assignment was randomized and balanced so that each task was evaluated under both workflows overall, while no participant encountered the same task twice.

We measured \emph{task success}, \emph{completion time}, \emph{expert-rated artifact quality}, and \emph{NASA-TLX}~\cite{hart88}. Success was defined as producing a structurally valid final pattern within the allotted time. To reduce rater burden, artifact quality was assessed on a stratified random sample. Five blinded experts took part in the evaluation, and each sampled artifact was independently rated by two experts on \emph{sewability}, \emph{structural balance}, and \emph{reference conformity}. We averaged these ratings into a composite score. NASA-TLX was administered immediately after each trial, enabling trial-level workload analysis. \re{To account for repeated observations, we used mixed-effects logistic regression for task success and linear mixed-effects models for completion time, artifact quality, and NASA-TLX.} After the study, we conducted a brief exit interview with each participant about their experience with the two workflows.

\subsection{Results}

\subsubsection{Performance benefits across task complexity}

Figure~\ref{fig:eval_performance} summarizes task success and completion time by workflow, difficulty, and expertise group. \re{TailorCoPilot significantly increased the odds of task success relative to the baseline ($OR = 2.31$, 95\% CI $[1.42, 3.86]$, $p = .001$).} The performance pattern differed by expertise. For novices, TailorCoPilot maintained a clear success-rate advantage across all difficulty levels, including simple tasks. For advanced novices, TailorCoPilot and the baseline were relatively close on simple tasks, but the gap widened on medium and complex tasks. This pattern suggests that TailorCoPilot was especially helpful when revision required multi-step coordination across related pattern pieces, while still providing substantial support for novices on simpler tasks.

\re{Completion times were significantly shorter with TailorCoPilot compared to the baseline ($p < .001$), with the effect size varied by task difficulty ($p = .018$).} TailorCoPilot was generally faster on medium and complex tasks in both groups. Among advanced novices, the CAD baseline remained competitive on some simple tasks, which is consistent with the efficiency of familiar tools when revision demands were still limited. Because the baselines differed by expertise group, we do not interpret absolute between-group differences as direct estimates of system effect.

\subsubsection{Revision became more efficient and manageable}

\begin{figure*}[!t]
    \centering
    \includegraphics[width=0.95\linewidth]{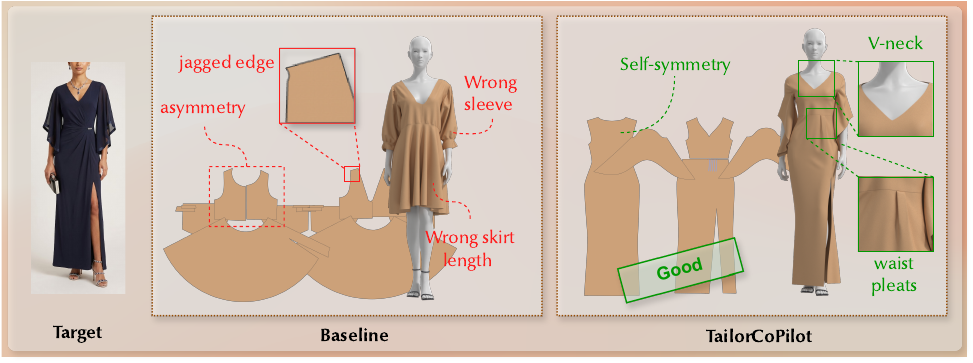}
    \caption{Representative qualitative result from the user study on the complex dress task. The baseline exhibited asymmetry, jagged edges, incorrect sleeve shape, and incorrect skirt length, while TailorCoPilot better preserved the target V-neck, waist pleats, and symmetry, producing a result that was structurally closer to the reference.}
    \label{fig:example}
    \Description{Representative qualitative result for a complex dress task, comparing the target image, a baseline result, and a TailorCoPilot result. The baseline shows asymmetry, jagged edges, incorrect sleeve shape, and incorrect skirt length, while TailorCoPilot better preserves symmetry, the V-neck, waist pleats, and the overall target structure.}
\end{figure*}

\re{TailorCoPilot not only improved artifact quality but also reduced perceived workload.} Table~\ref{tab:expert_scores} reports blinded expert ratings of final artifacts, averaged across task difficulties. \re{These ratings demonstrated strong inter-rater reliability ($ICC(1,1) = 0.81$). Artifact quality was significantly higher with TailorCoPilot ($p = .001$).} The clearest gains appeared in \emph{structural balance} and \emph{reference conformity}, suggesting better preservation of cross-piece consistency and closer alignment with the target design. Gains in \emph{sewability} were also consistent, though relatively smaller.
Representative qualitative results further illustrate this pattern (Figure~\ref{fig:example}). In the complex dress task, the baseline showed asymmetry, jagged edges, incorrect sleeve shape, and incorrect skirt length, while TailorCoPilot better preserved the target V-neck, waist pleats, and symmetry, producing a result that was structurally closer to the reference.

\re{Perceived workload was also reduced with TailorCoPilot, as measured by NASA-TLX scores ($p < .001$). Exit interviews further indicated that participants retained a sense of agency and ownership during revision. One participant described the result as ``still my pattern because the final decisions were mine'' (PU6), while an advanced novice remarked, ``I stopped second-guessing every edit'' (PU13). These findings indicate that TailorCoPilot made pattern revision more manageable for less experienced users without diminishing their sense of control.}
% NASA-TLX scores were likewise lower under TailorCoPilot, with larger reductions relative to the baseline observed for medium and complex tasks. Together, these findings suggest that TailorCoPilot made pattern making more approachable and manageable for less experienced users. As one advanced novice put it, ``I stopped second-guessing every edit'' (PU13).

\begin{table*}[t]
\centering
\caption{\textbf{Expert ratings of final artifacts} (\textbf{mean $\pm$ SD}) on a \textbf{1--5} scale, grouped by user expertise. Values are averaged across simple, medium, and complex tasks. Across both expertise groups, TailorCoPilot received higher ratings than the baseline on all evaluation dimensions.}
\label{tab:expert_scores}
\small
\renewcommand{\arraystretch}{1.15}
\setlength{\tabcolsep}{5pt}
\begin{tabular}{lcccccccc}
\toprule
& \multicolumn{2}{c}{\textbf{Sewability}}
& \multicolumn{2}{c}{\textbf{Structural balance}}
& \multicolumn{2}{c}{\textbf{Reference conformity}}
& \multicolumn{2}{c}{\textbf{Overall quality}} \\
\cmidrule(lr){2-3}\cmidrule(lr){4-5}\cmidrule(lr){6-7}\cmidrule(lr){8-9}
\textbf{Expertise}
& \textbf{TailorCoPilot} & \textbf{Baseline}
& \textbf{TailorCoPilot} & \textbf{Baseline}
& \textbf{TailorCoPilot} & \textbf{Baseline}
& \textbf{TailorCoPilot} & \textbf{Baseline} \\
\midrule
Novice
& \cellcolor{brandred!63} 3.82 $\pm$ 0.86
& \cellcolor{brandred!21} 2.02 $\pm$ 1.01
& \cellcolor{brandred!66} 3.97 $\pm$ 0.81
& \cellcolor{brandred!8} 1.48 $\pm$ 0.59
& \cellcolor{brandred!66} 3.97 $\pm$ 0.83
& \cellcolor{brandred!5} 1.35 $\pm$ 0.53
& \cellcolor{brandred!65} 3.92 $\pm$ 0.82
& \cellcolor{brandred!11} 1.62 $\pm$ 0.70 \\

Advanced novice
& \cellcolor{brandred!71} 4.17 $\pm$ 0.74
& \cellcolor{brandred!55} 3.48 $\pm$ 0.93
& \cellcolor{brandred!75} 4.32 $\pm$ 0.70
& \cellcolor{brandred!32} 2.48 $\pm$ 0.99
& \cellcolor{brandred!75} 4.33 $\pm$ 0.71
& \cellcolor{brandred!35} 2.63 $\pm$ 0.97
& \cellcolor{brandred!74} 4.27 $\pm$ 0.71
& \cellcolor{brandred!41} 2.87 $\pm$ 0.95 \\
\bottomrule
\end{tabular}
\Description{Expert ratings of final artifacts on a 1 to 5 scale, grouped by user expertise. TailorCoPilot receives higher ratings than the baseline on sewability, structural balance, reference conformity, and overall quality for both novices and advanced novices. For novices, TailorCoPilot scores 3.82, 3.97, 3.97, and 3.92, compared with 2.02, 1.48, 1.35, and 1.62 for the baseline. For advanced novices, TailorCoPilot scores 4.17, 4.32, 4.33, and 4.27, compared with 3.48, 2.48, 2.63, and 2.87 for the baseline. Standard deviations range from 0.70 to 0.93 for TailorCoPilot and from 0.53 to 1.01 for the baseline, indicating that TailorCoPilot maintains stronger performance across all evaluation dimensions in both expertise groups.}
\end{table*}

\subsection{Diagnostic Study on TailorTrace}

% We compared two variants on a set of complex tasks. (\textbf{1}) \textbf{TailorCoPilot (T)}\footnote{T refers to Textbook, while F refers to Full.}, which used traces extracted from textbooks but lacked expert-authored production traces. (\textbf{2}) \textbf{TailorCoPilot (F)}, which additionally incorporated expert-authored traces of industrial pattern-making practices. To isolate the impact of expert data, both variants shared the same target references, base patterns, and inference budgets across all tasks.

% Across representative cases, \textbf{TailorCoPilot (T)} often produced revisions in which individual piece adjustments moved in the right direction, but the coordinated structural relations across pieces were left incomplete. This suggests that static construction knowledge can support locally plausible edits, yet provides limited guidance for coordinating interdependent multi-piece revisions. By contrast, \textbf{TailorCoPilot (F)} more reliably produced structurally valid and reference-consistent final patterns, as shown in Figure~\ref{fig:ablation}. These results suggest that TailorTrace contributes more than static garment-construction knowledge alone, especially for revisions that require coordinated changes across multiple pattern pieces.

\re{To systematically examine the contribution of trace-based scaffolding, we expanded the diagnostic study to 15 complex pattern-making tasks and compared three conditions. \textbf{TailorCoPilot (V)} relied solely on the pre-trained Qwen3-VL-8B, without access to TailorTrace data. \textbf{TailorCoPilot (T)} and \textbf{TailorCoPilot (F)} both utilized textbook-based traces, with the latter also incorporating expert-authored traces capturing industrial pattern-making practices. All conditions used identical target references, base patterns, and inference budgets, and all outputs were expert-evaluated using the same protocol as in the main study.
}

\re{The results showed that \textbf{V} achieved no success across the 15 tasks and received an average expert rating of $M=1.3$ ($SD=0.5$). \textbf{T} met the success criterion in 3 tasks (20.0\%) with an average rating of $M=1.9$ ($SD=0.8$). In contrast, \textbf{F} met the criterion in 10 tasks (66.7\%) and scored significantly higher on expert evaluation ($M=4.1$, $SD=0.6$). A paired Wilcoxon signed-rank test confirmed a statistically significant difference between \textbf{T} and \textbf{F} ($W=1$, $p<.001$).
}

\begin{figure}[!t]
    \centering
    \includegraphics[width=0.8\linewidth]{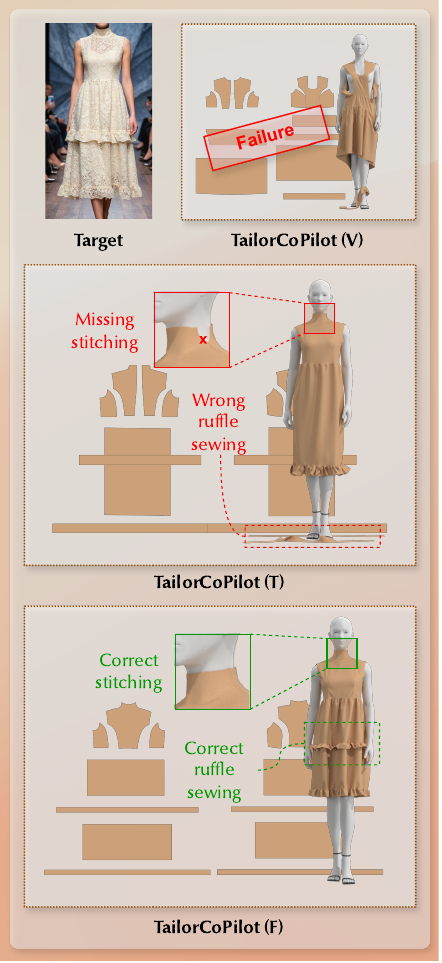}
    \caption{\re{Qualitative comparison from the diagnostic study using identical references, base patterns, and inference budgets. \textbf{TailorCoPilot (V)} fails to execute the revision. \textbf{TailorCoPilot (T)} produces locally plausible edits but breaks cross-piece structural relations (e.g., missing stitching and incorrect ruffle construction). By incorporating expert traces, \textbf{TailorCoPilot (F)} generates a structurally valid pattern that accurately matches the target design.}}
    \label{fig:ablation}
    \Description{Diagnostic comparison on a representative complex task. The VLM-only version (V) fails to complete the requested revision. The textbook-only version (T) produces locally plausible edits but leaves cross-piece relations incomplete, showing missing stitching and incorrect ruffle construction. The full version (F) with expert-authored traces produces correct stitching and ruffle construction, closely matching the target reference.}
\end{figure}

\re{Figure~\ref{fig:ablation} illustrates a representative qualitative example of these differences. While \textbf{V} completely failed the revision task and \textbf{T} only managed localized modifications without maintaining overall structural integrity, \textbf{F} successfully generated a reference-consistent pattern. These results indicate that expert-authored TailorTrace data offers critical procedural guidance that extends beyond static LLM knowledge and textbook materials, particularly for complex tasks requiring coordinated changes across multiple pattern components.}

\section{Discussion}

\begin{figure}[!t]
  \centering
  \includegraphics[width=\linewidth]{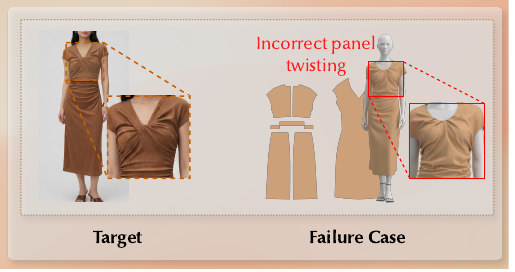}
  \caption{Failure Case. As TailorTrace and TailorCoPilot are primarily defined by 2D topological transformations, they currently struggle to model intricate physical manipulations, such as the twisted knot on the target bust.}
  \label{fig:discussion}
  \Description{Failure case showing a target dress with a twisted knot detail at the bust and the system’s incorrect output. Because the method mainly models 2D topological transformations, it struggles to represent intricate physical manipulations and produces incorrect panel distortion instead of the intended twist.}
\end{figure}

\subsection{Reusable Traces as Expert Process Knowledge}

Our results suggest that one central contribution of TailorCoPilot lies in making expert process knowledge usable during pattern making. In garment pattern making, difficulty often comes from deciding how design intent should be translated into concrete edits, how a change to one piece propagates to related pieces, and how to continue when an intermediate state is only partially correct. TailorTrace addresses these challenges by representing pattern development as discrete states linked by explicit geometric operations. This gives TailorCoPilot a process-aware basis for planning, revision, and recovery.

This also clarifies what agentic pattern making means in our setting. TailorCoPilot does more than generate a final result from a prompt. It operates over traceable states, proposes multi-step revisions, and supports users in inspecting and adjusting the process as it unfolds. Its agentic behavior therefore depends on a structured revision space that remains visible and editable to the user.

Our findings also show why static construction knowledge alone is often insufficient. Handbooks can describe principles and canonical techniques, but they do not fully preserve how experts sequence edits, maintain cross-piece consistency, or carry partial results forward through complex revision. The diagnostic study points to this gap directly. In the textbook-only variant, the system could still produce locally plausible edits, but it was less reliable at coordinating structurally valid changes across pieces. This suggests that expert-authored traces contribute procedural information that is difficult to recover from rules or final examples alone. It also provides a foundation for collecting process-rich data that may support future generative AI models in this domain.

\subsection{Supporting Apprenticeship Without Replacing Expertise}

TailorCoPilot supported novices and advanced novices in related but not identical ways. For novices, it made the work more approachable by reducing the amount of CAD fluency and pattern-making knowledge required to begin making meaningful changes. For advanced novices, its value was less about entry and more about continuity. Instead of having to reconstruct expert procedures step by step, they could follow a more legible path through the revision process, with key operations, dependencies, and constraints surfaced at the moment they mattered. Across both groups, this made complex revision easier to start and easier to continue.

At the same time, these gains did not erase the gap between assisted participation and expert practice. Pattern making still relies on forms of judgment that resist full formalization, such as evaluating balance, assessing silhouette quality, and deciding whether a local correction preserves the coherence of the overall construction. What shifted was not the need for expertise itself, but the distribution of effort. TailorCoPilot reduced procedural overhead and made recovery from mistakes less demanding, while higher-level evaluation and final decisions remained in the user's hands.

This suggests a broader role for systems of this kind in experience-dependent workflows shaped by iteration and expert judgment. The aim is not to flatten the distinction between novice and professional performance, but to make expert process knowledge more available during learning. Seen this way, TailorCoPilot is better understood as support for apprenticeship than as a substitute for expertise. By preserving senior experts' workflows as reusable traces, it points to one possible way of addressing the generational skills gap that motivates this work.

\subsection{Limitations and Future Directions}

This work has three main limitations. First, the evaluation uses a modest sample and measures short-term performance in a controlled setting. Although the results suggest that trace-based support improves revision performance and usability, we do not yet know whether these benefits translate into durable learning or long-term skill development. Longitudinal studies will be needed to examine whether repeated use supports independent pattern-making ability over time.

Second, TailorCoPilot depends on a scoped representation of the domain. The current system assumes a predefined operator vocabulary, and expert-authored traces that can be represented through TailorTrace. \re{Given the modest dataset of 93 traces and approximately 2,500 derived training pairs, our claims are limited to operation synthesis on held-out traces within this bounded garment domain; broader deployment in professional prototyping and small-batch production remains to be validated.} Figure~\ref{fig:discussion} shows a representative failure case, where the system only partially recovered the intended structure under these constraints.

Third, to ensure comparability across conditions, we used a unified set of default fabric-simulation parameters. This simplifies evaluation, but leaves material-specific behavior and some forms of production variability for future work. Extending TailorTrace to richer material settings and broader garment categories remains an important direction.

\section{Conclusion}

This paper presented \textbf{TailorCoPilot}, a system that supports garment pattern making by grounding interaction in \textbf{TailorTrace}, a version-controlled representation of pattern states and transformations. Our findings show that representing expert process knowledge as reusable traces can make pattern making more accessible to novices and advanced novices, especially in revisions that require coordination across multiple pieces and intermediate states. More broadly, this work suggests that practice-based expertise may become more usable when it is captured as revisitable traces. Within this bounded garment-pattern domain, our findings point to a promising direction for supporting both apprenticeship and future AI systems in workflows where expert knowledge is deeply procedural and difficult to formalize.
\begin{acks}
The authors gratefully acknowledge the support provided by Style3D Research. The authors would also like to acknowledge the financial support from the Fundamental Research Funds for the Central Universities (Grant Nos. CUSF-DH-T-2025007 and 2232026G-08) and the International Cooperation Fund of the Science and Technology Commission of Shanghai Municipality (Grant No. 21130750100).
\end{acks}

\bibliographystyle{ACM-Reference-Format}
\bibliography{sample-base}

% %%
% %% If your work has an appendix, this is the place to put it.
% \appendix









\end{document}

\title{TailorCoPilot: Enabling Agentic Pattern Making with Version-Controlled State Tracking \\ (Supplementary Material)}

\maketitle

\section{Study Materials}

\subsection{Participant Information}

Tables~\ref{tab:participant_details_formative} and~\ref{tab:participant_details_user} report supplementary participant information for the formative study and user study, including demographics, years of pattern-making experience (Exp.), tool accessibility in the study, and compensation. Following the grouping used in the main paper, participants with less than 0.5 years of pattern-making experience were categorized as novices, and those with 0.5 to 2 years of experience were categorized as advanced novices. Experts were senior pattern makers or educators with more than 5 years of industry experience. The AI column refers to whether AI-based pattern assistants~\cite{Zhou24,Nakayama24} are used in the study, and Style3D~\cite{Style3D} is a professional garment CAD and 3D draping system. Checkmarks indicate that participants were able to operate each tool after a simple 10-minute training session\footnote{Participants in the formative study received \$20, and participants in the user study received \$30.}.

\begin{table}[t]
\centering
\footnotesize
\setlength{\tabcolsep}{3.5pt}
\renewcommand{\arraystretch}{1.05}
\caption{Participant information for the formative study.}
\label{tab:participant_details_formative}
\begin{tabular}{lcccccc}
\toprule
ID & Group & Age & Gender & Exp. & AI & Style3D \\
\midrule
PF1  & Novice          & 19 & F & 0.0  & \checkmark & \\
PF2  & Novice          & 20 & F & 0.0  & \checkmark & \\
PF3  & Novice          & 20 & F & 0.0  & \checkmark & \\
PF4  & Novice          & 21 & M & 0.0  & \checkmark & \\
PF5  & Novice          & 21 & F & 0.2  & \checkmark & \\
PF6  & Novice          & 22 & F & 0.0  & \checkmark & \\
PF7  & Advanced novice & 19 & F & 0.5  & \checkmark & \\
PF8  & Advanced novice & 20 & F & 1.0  & \checkmark & \\
PF9  & Advanced novice & 21 & M & 1.5  & \checkmark & \checkmark \\
PF10 & Advanced novice & 22 & F & 2.0  & \checkmark & \checkmark \\
PF11 & Expert          & 36 & F & 8.0  & \checkmark & \checkmark \\
PF12 & Expert          & 41 & M & 10.5 & \checkmark & \checkmark \\
PF13 & Expert          & 45 & F & 14.0 & \checkmark & \\
\bottomrule
\end{tabular}
\Description{Participant information for the formative study. The table lists 13 participants grouped as novice, advanced novice, and expert. Columns report participant ID, age, gender, years of pattern-making experience, and whether each participant could use AI assistants and Style3D after brief training. Ages range from 19 to 45 years, and experience ranges from 0.0 to 14.0 years.}
\end{table}

\begin{table}[t]
\centering
\footnotesize
\setlength{\tabcolsep}{3.5pt}
\renewcommand{\arraystretch}{1.05}
\caption{Participant information for the user study.}
\label{tab:participant_details_user}
\begin{tabular}{lcccccc}
\toprule
ID & Group & Age & Gender & Exp. & AI & Style3D \\
\midrule
PU1  & Novice          & 19 & F & 0.0 & \checkmark & \\
PU2  & Novice          & 19 & F & 0.0 & \checkmark & \\
PU3  & Novice          & 20 & M & 0.0 & \checkmark & \\
PU4  & Novice          & 20 & F & 0.0 & \checkmark & \\
PU5  & Novice          & 20 & F & 0.2 & \checkmark & \\
PU6  & Novice          & 21 & F & 0.0 & \checkmark & \\
PU7  & Novice          & 21 & F & 0.0 & \checkmark & \\
PU8  & Novice          & 21 & M & 0.3 & \checkmark & \\
PU9  & Novice          & 22 & F & 0.0 & \checkmark & \\
PU10 & Novice          & 22 & F & 0.0 & \checkmark & \\
PU11 & Advanced novice & 19 & F & 0.5 & \checkmark & \\
PU12 & Advanced novice & 19 & M & 0.8 & \checkmark & \checkmark \\
PU13 & Advanced novice & 20 & F & 1.0 & \checkmark & \checkmark \\
PU14 & Advanced novice & 20 & F & 1.2 & \checkmark & \\
PU15 & Advanced novice & 21 & F & 1.4 & \checkmark & \checkmark \\
PU16 & Advanced novice & 21 & M & 1.6 & \checkmark & \checkmark \\
PU17 & Advanced novice & 21 & F & 1.7 & \checkmark & \checkmark \\
PU18 & Advanced novice & 22 & F & 1.8 & \checkmark & \\
PU19 & Advanced novice & 22 & M & 1.9 & \checkmark & \checkmark \\
PU20 & Advanced novice & 22 & F & 2.0 & \checkmark & \checkmark \\
\bottomrule
\end{tabular}
\Description{Participant information for the user study. The table lists 20 participants grouped as novice and advanced novice. Columns report participant ID, age, gender, years of pattern-making experience, and whether each participant could use AI assistants and Style3D after brief training. Ages range from 19 to 22 years, and experience ranges from 0.0 to 2.0 years.}
\end{table}

\subsection{Formative Study Interview Protocol}
\label{app:formative_interview}

This section summarizes the semi-structured interview protocol used in the formative study. The protocol covered the following domains:

\begin{enumerate}
    \item \textbf{Current Workflow and Tools}

    \begin{quote}
    ``Can you walk me through your typical workflow when developing a new pattern?''
    
    ``Which tools, references, or examples do you usually rely on during this process?''
    \end{quote}

    \item \textbf{Learning and Knowledge Acquisition}

    \begin{quote}
    ``How did you learn pattern making, and what helped you most?''
    
    ``Which aspects are difficult to learn or teach?''
    \end{quote}

    \item \textbf{Decomposition of Garment Changes}

    \begin{quote}
    ``When you want to change a garment design, how do you decompose it into concrete pattern edits?''
    
    ``How do you decide what to modify first and what should follow next?''
    \end{quote}

    \item \textbf{Breakdowns and Repair Strategies}

    \begin{quote}
    ``What are the most challenging parts of pattern making for you?''
    
    ``How do you usually recover when a modification does not work as expected?''
    \end{quote}

    \item \textbf{Judgment Criteria}

    \begin{quote}
    ``How do you judge whether a modification is structurally sound?''
    
    ``How do you determine whether the resulting garment still matches the intended appearance or style?''
    \end{quote}
\end{enumerate}

\noindent\textit{Note.} Representative questions are provided for illustration. Follow-up questions were used to probe specific examples, decision points, and repair strategies raised by participants.

\subsection{Task Inventory}

Below we list the full task pool used in the study. AI assistants are primarily used for pattern initialization, while participants were allowed to use Style3D to adjust and refine the base pattern directly.

\subsubsection*{Simple Tasks (15)}
\begin{enumerate}[label=(\arabic*),leftmargin=2em,labelsep=0.5em,itemsep=1pt,topsep=2pt]
    \item Shorten the bodice
    \item Lengthen the bodice
    \item Widen the waist
    \item Widen the hem
    \item Broaden the shoulder width
    \item Deepen the back neckline
    \item Increase the neckline width
    \item Increase collar stand height
    \item Widen the sleeve opening
    \item Lengthen the skirt
    \item Shorten the skirt
    \item Increase the pants waist circumference
    \item Widen the pant leg opening
    \item Increase the skirt flare
    \item Lengthen the sleeve
\end{enumerate}

\subsubsection*{Medium Tasks (10)}
\begin{enumerate}[label=(\arabic*),leftmargin=2em,labelsep=0.5em,itemsep=1pt,topsep=2pt]
    \item Raise the armhole and adjust sleeve cap matching
    \item Transfer a waist dart to the side seam
    \item Transfer a bust dart to a princess seam
    \item Add pleats to the front panel and reconfigure the waist dart arrangement
    \item Add layered structure to the skirt and adjust the flare volume
    \item Modify the collar type, seam lines, and hems
    \item Convert a standard sleeve into a dropped sleeve
    \item Convert a straight sleeve into a puff sleeve
    \item Add pleats to the knees and widen the leg hems
    \item Add a center-front placket and restructure the front panel
\end{enumerate}

\subsubsection*{Complex Tasks (5)}
\begin{enumerate}[label=(\arabic*),leftmargin=2em,labelsep=0.5em,itemsep=1pt,topsep=2pt]
    \item Create a dress pattern from a reference image (fashion sketch)
    \item Create a dress pattern from a reference image (garment photograph)
    \item Create a coat pattern from a reference image (fashion sketch)
    \item Create a coat pattern from a reference image (garment photograph)
    \item Create a dress pattern from a text description
\end{enumerate}

\section{System Details}

\subsection{Interface Overview}
\begin{figure*}[t]
\centering
\includegraphics[width=\linewidth]{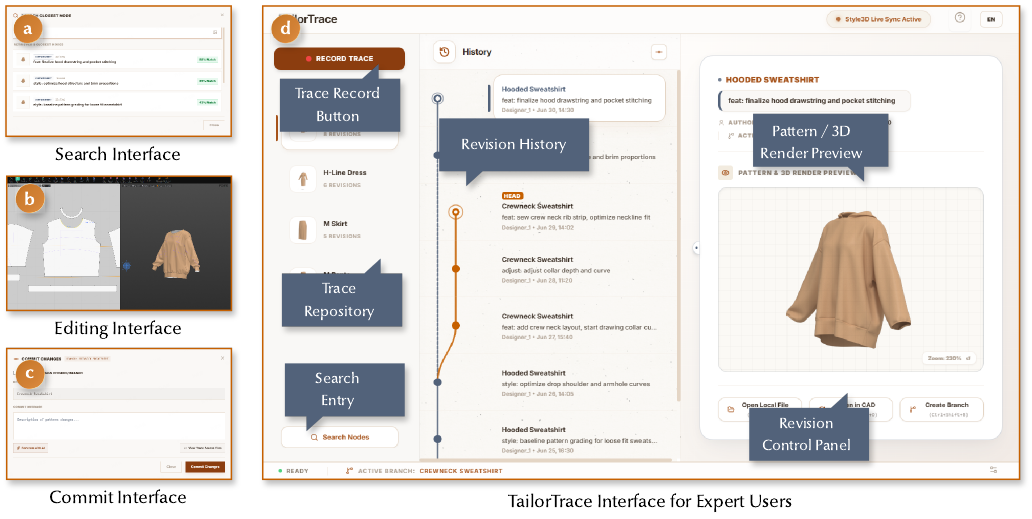}
\caption{\re{Overview of the TailorTrace interface for expert users. (a) Search interface for locating entries in the trace repository. (b) Editing interface for exploring garment designs using CAD software while recording operation traces. (c) Commit interface for submitting the current design state. (d) Main TailorTrace interface, comprising trace records, a search entry, revision history, 2D pattern and 3D render previews, and a revision control panel.}}
\label{fig:app-ui-0}
\Description{The TailorTrace interface for expert users consists of four views. The search interface allows users to locate entries in the trace repository. The editing interface supports garment design exploration in CAD software while recording the user's operations as traces. The commit interface allows users to submit the current design state. The main expert-facing interface integrates trace records, an entry point to the search interface, revision history, 2D pattern and 3D render previews, and a revision control panel for inspecting and navigating design revisions.}
\end{figure*}

\begin{figure*}[t]
\centering
\includegraphics[width=\linewidth]{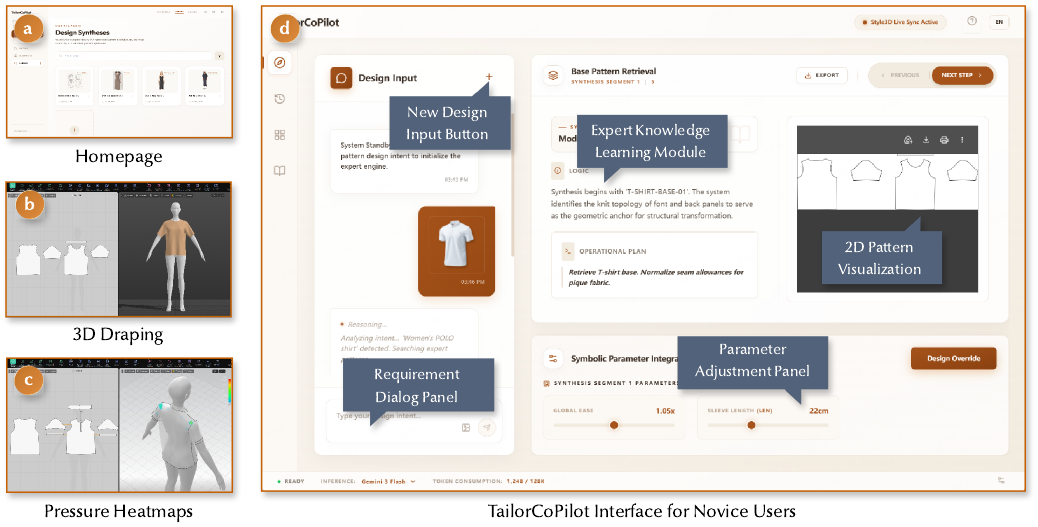}
\caption{TailorCoPilot interface overview. (a) Homepage of TailorCoPilot. (b) 3D draping workspace for inspecting garment geometry. (c) Pressure heatmap view for assessing local fit and tension. (d) Main interface for novice users, integrating a requirement dialog panel, an expert knowledge learning module, 2D pattern visualization, and a parameter adjustment panel.}
\label{fig:app-ui-1}
\Description{Composite interface overview of TailorCoPilot across four views. The figure shows the homepage, a 3D draping workspace, a pressure heatmap view, and a main novice-facing editing interface that combines requirement input, expert guidance, 2D pattern visualization, and parameter adjustment for iterative pattern editing.}
\end{figure*}

\re{In TailorTrace (Figure~\ref{fig:app-ui-0}), experts edit garment patterns using native CAD operations while the system records the resulting semantic operations as traces in the background. They can search existing traces in the repository, commit intermediate design states, create branches from selected revisions, and reload previous versions into the CAD workspace for further editing. Each committed state is validated and stored alongside its 2D pattern geometry, 3D preview, and captured operation sequence, displayed side by side to support revision comparison and reuse. The collected expert traces then serve as training data for TailorCoPilot.

In TailorCoPilot (Figure~\ref{fig:app-ui-1}), novices specify their design requirements through the dialog panel and receive an expert-guided adjustment plan. They can apply or modify symbolic parameters while inspecting the updated 2D patterns, and use the 3D draping and pressure heatmap views to evaluate the resulting geometry and fit. This supports an iterative workflow of requirement specification, pattern adjustment, and result evaluation.
}

\subsection{Pattern Representation}

\begin{figure}[ht]
    \centering
    \includegraphics[width=0.8\linewidth]{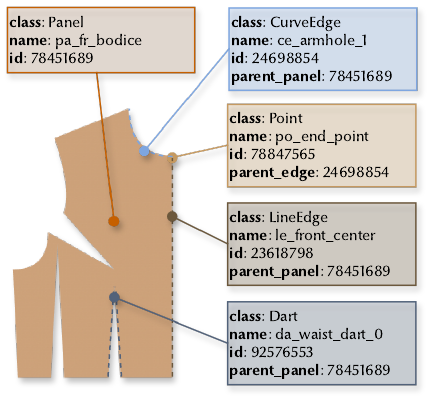}
    \caption{Structured sewing-pattern state representation. Each pattern is encoded as typed entities, including panels, edges, points, and darts, linked by unique identifiers and parent-child relations.}
    \Description{Diagram of a structured sewing-pattern representation. A bodice front panel is shown in the center, with callouts labeling five typed entities: one panel, one curved edge, one straight edge, one point, and one dart. Each entity has a unique identifier. The curved edge, straight edge, and dart reference the same parent panel, while the point references the curved edge as its parent. The figure illustrates how pattern elements are represented as typed objects linked by parent-child relationships.}
    \label{fig:pattern-representation}
\end{figure}

Figure~\ref{fig:pattern-representation} illustrates the structured sewing-pattern representation used in TailorCoPilot. Each pattern is represented as a set of typed entities, such as panels, edges, points, and darts, each associated with a unique identifier and linked through parent-child references. This representation provides the structural schema on which symbolic editing operations and state tracking are defined.

\subsection{TailorTrace Log Example}

\begin{figure}[ht]
    \centering
    \includegraphics[width=\linewidth]{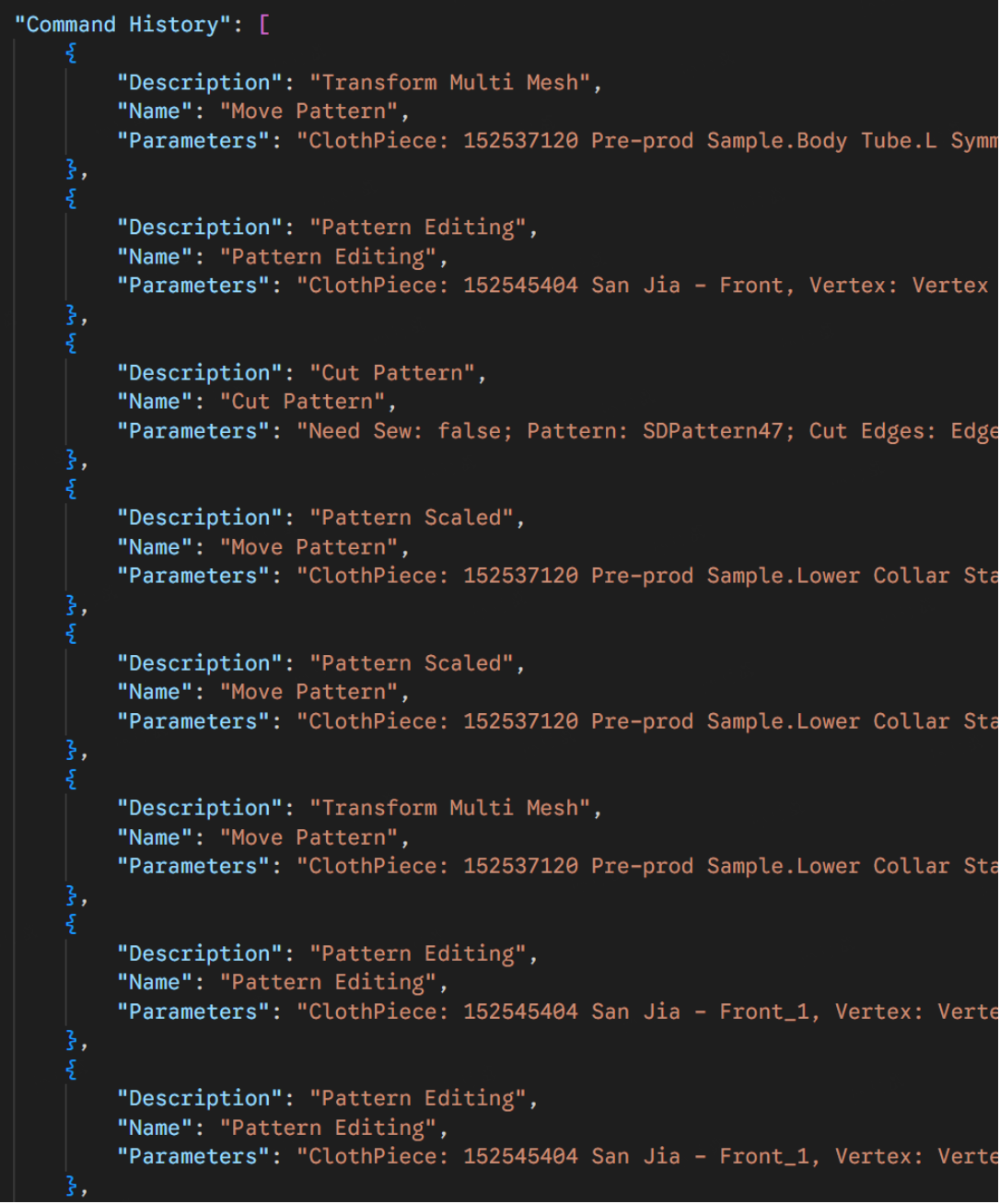}
    \caption{Representative excerpt of the raw TailorTrace log format.}
    \Description{Screenshot of a raw TailorTrace log excerpt in a code-like JSON format. The entries record low-level editing history from the pattern-making environment, including action descriptions, operation names, and associated parameter fields.}
    \label{fig:app-history}
\end{figure}

Figure~\ref{fig:app-history} presents a representative excerpt of the raw TailorTrace log, illustrating how editing actions are recorded during pattern making. The log preserves low-level operation records exported from the underlying environment.

\subsection{Symbolic Operators}
\begin{figure}[t]
\centering
\includegraphics[width=\linewidth]{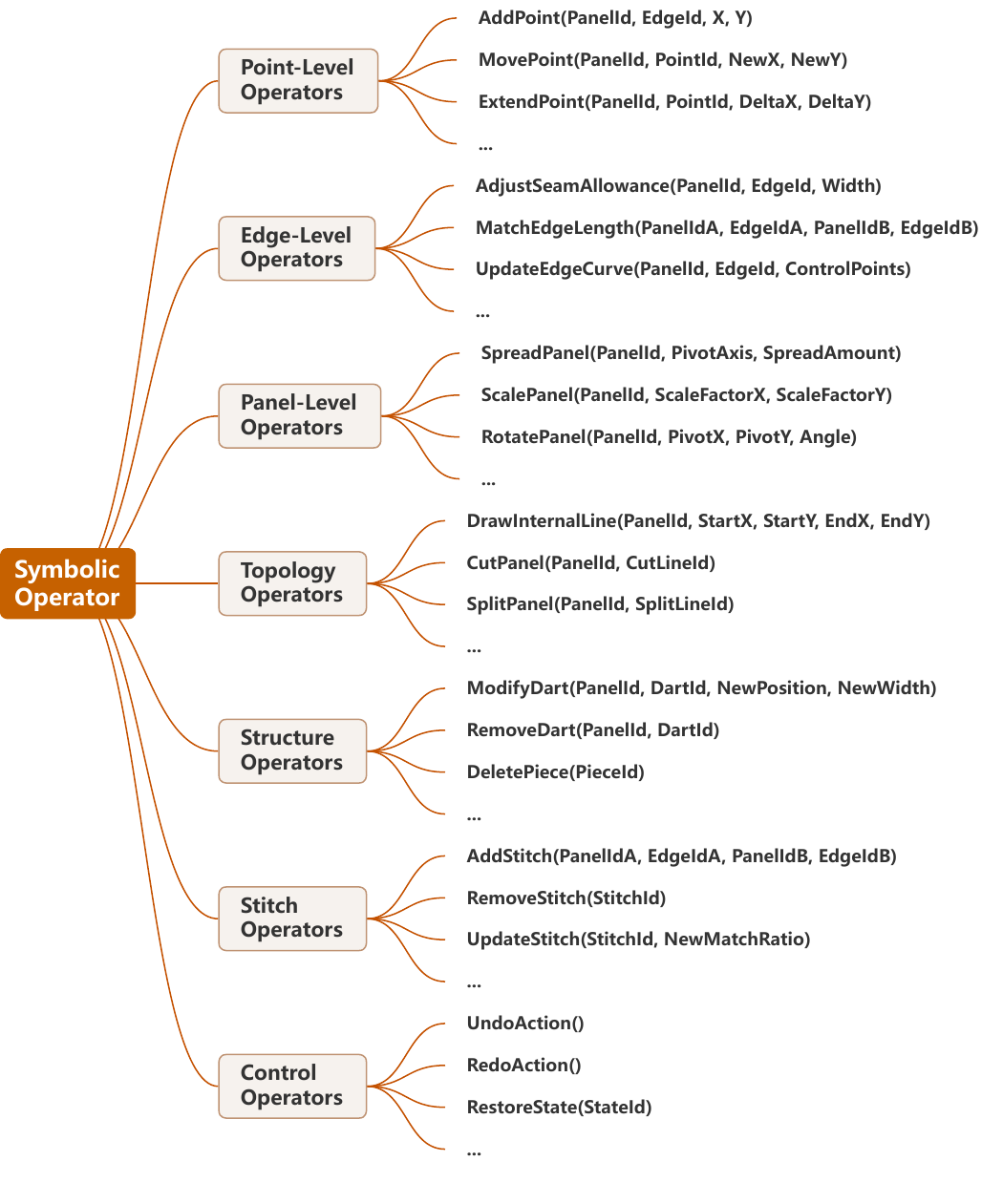}
\caption{Representative symbolic operators used in TailorTrace and TailorCoPilot. The operator set provides the executable transformation layer that maps symbolic editing plans to concrete pattern-state revisions.}
\label{fig:operators}
\Description{Diagram of representative symbolic operators used in TailorTrace and TailorCoPilot. A central symbolic-operator node branches to seven groups: point-level, edge-level, panel-level, topology, structure, stitch, and control operators. Each group includes example commands that represent concrete pattern-editing actions.}
\end{figure}

This appendix presents representative symbolic operators implemented in TailorTrace and used by TailorCoPilot. These operators illustrate the executable action space through which high-level reasoning outputs are compiled into concrete transformations over sewing-pattern states. It covers local geometric edits, edge and stitch operations, piece-structure modifications, topological restructuring, and history-aware control, thereby providing the operational interface between symbolic plans and editable pattern outcomes, as shown in Figure~\ref{fig:operators}.

\subsection{Base Pattern Retrieval}
\re{The retrieval library contains 120 editable base garments across 6 coarse categories, including tops, dresses, skirts, trousers, jackets, and coats. Each item consists of a validated sewing-pattern state, front-view rendering, garment metadata, and category label. Library patterns were internally authored and manually checked to ensure that they were editable, sewable, and compatible with the shared simulation setup.} 

We use RADIO~\cite{Ranzinger_2024_CVPR} as the visual retriever for pattern-library initialization. 
All library garments are rendered under a fixed retrieval setup with a shared avatar, camera configuration, background, and lighting, so that nearest-neighbor search is driven primarily by garment silhouette and local design cues rather than rendering variation. 
Each library item is indexed with front view, and retrieval is performed by cosine similarity in the RADIO embedding space. 
At inference time, the system retrieves the top-4 candidates and uses the highest-ranked valid item as the initial editable base pattern. 
For text-only inputs, retrieval is first restricted to a coarse garment category predicted from the user request, after which RADIO is applied within the filtered subset. 
When the similarity score falls below a preset threshold $\tau_r$, the interface returns the top-4 candidates for user selection.

\subsection{Cloth Simulation Settings}

To ensure consistency across experiments, all simulations use a shared set of cloth parameters, summarized in Table~\ref{tab:cloth_params}.

\begin{table}[h]
\centering
\footnotesize
\caption{Cloth simulation parameters used across experiments.}
\label{tab:cloth_params}
\begin{tabular}{lll}
\toprule
\textbf{Category} & \textbf{Parameter} & \textbf{Value} \\
\midrule
Stretch & Stretch Stiffness & 15 \\
Bend & Bend Stiffness & 20 \\
General & Areal Density (GSM) & 66.67 \\
General & Thickness (mm) & 0.14 \\
\midrule
Advanced & Bend Damping Ratio & 0 \\
Advanced & Bend Buckling Ratio & 0 \\
Advanced & Dynamic Friction Coefficient & 0.03 \\
Advanced & Static Friction Coefficient & 0.03 \\
\midrule
Structural & Particle Distance (mm) & 6.00 \\
Structural & Layers & 0 \\
Structural & Warp Density (\%) & 100.00 \\
Structural & Weft Density (\%) & 100.00 \\
Structural & Additional Thickness (mm) & 0.00 \\
Structural & Collision Thickness (mm) & 1.50 \\
\bottomrule
\end{tabular}
\Description{Cloth simulation parameters used across experiments. The table groups parameters into stretch, bend, general, advanced, and structural settings, including stiffness, areal density, thickness, friction coefficients, particle distance, fabric density, and collision thickness. Representative values include stretch stiffness 15, bend stiffness 20, areal density 66.67 GSM, thickness 0.14 mm, particle distance 6.00 mm, and collision thickness 1.50 mm.}
\end{table}

\subsection{Prompting Architecture}
We implement TailorCoPilot as a modular prompting architecture with progressive disclosure. Instead of presenting all instructions, operator definitions, and repair heuristics in a single monolithic prompt, the system maintains a compact core prompt and conditionally attaches stage-specific modules only when they are needed at inference time. This design reduces prompt bloat while preserving stage-specific precision and keeping model outputs aligned with executable symbolic operators.

\paragraph{Core prompt.}
The core prompt provides the persistent guidance shared across all stages, including the model role, the structured sewing-pattern state schema, general validity constraints, and the high-level policy for symbolic operator usage.

\paragraph{Planning module.}
When the system translates a target intent and the current pattern state into an executable operator sequence, it activates a planning prompt that requests a stepwise plan grounded in the current state, the available operator vocabulary, and optional retrieved references. This module asks the model to identify the main discrepancies between the current state and the target intent, decompose the transformation into interpretable local edits, preserve cross-piece consistency, and make stitch updates explicit when needed.

\paragraph{Referenced resources.}
To keep the shared prompt context compact, detailed operator signatures, state-field descriptions, formatting constraints, and representative examples are maintained as lightweight referenced resources and introduced only when required by the active stage.

Overall, this modular design supports a planning, validation, and repair loop while avoiding unnecessary duplication across prompts.

\subsection{TailorCoPilot Dataset Generation}
\re{The TailorCoPilot dataset was constructed from validated TailorTrace state transitions after applying a trace-level train/validation split. For each sampled source-target pair, Gemini was used to draft a natural-language design-intent description, which domain experts then manually verified and corrected for technical terminology and consistency with the actual edit. The final dataset contains approximately 2,500 description-operation pairs and is used to evaluate operation synthesis on held-out traces, rather than large-scale zero-shot generalization.}

\section{Extended Results}

\subsection{Additional Statistical Results}

\begin{figure*}[t]
\centering
\includegraphics[width=0.9\linewidth]{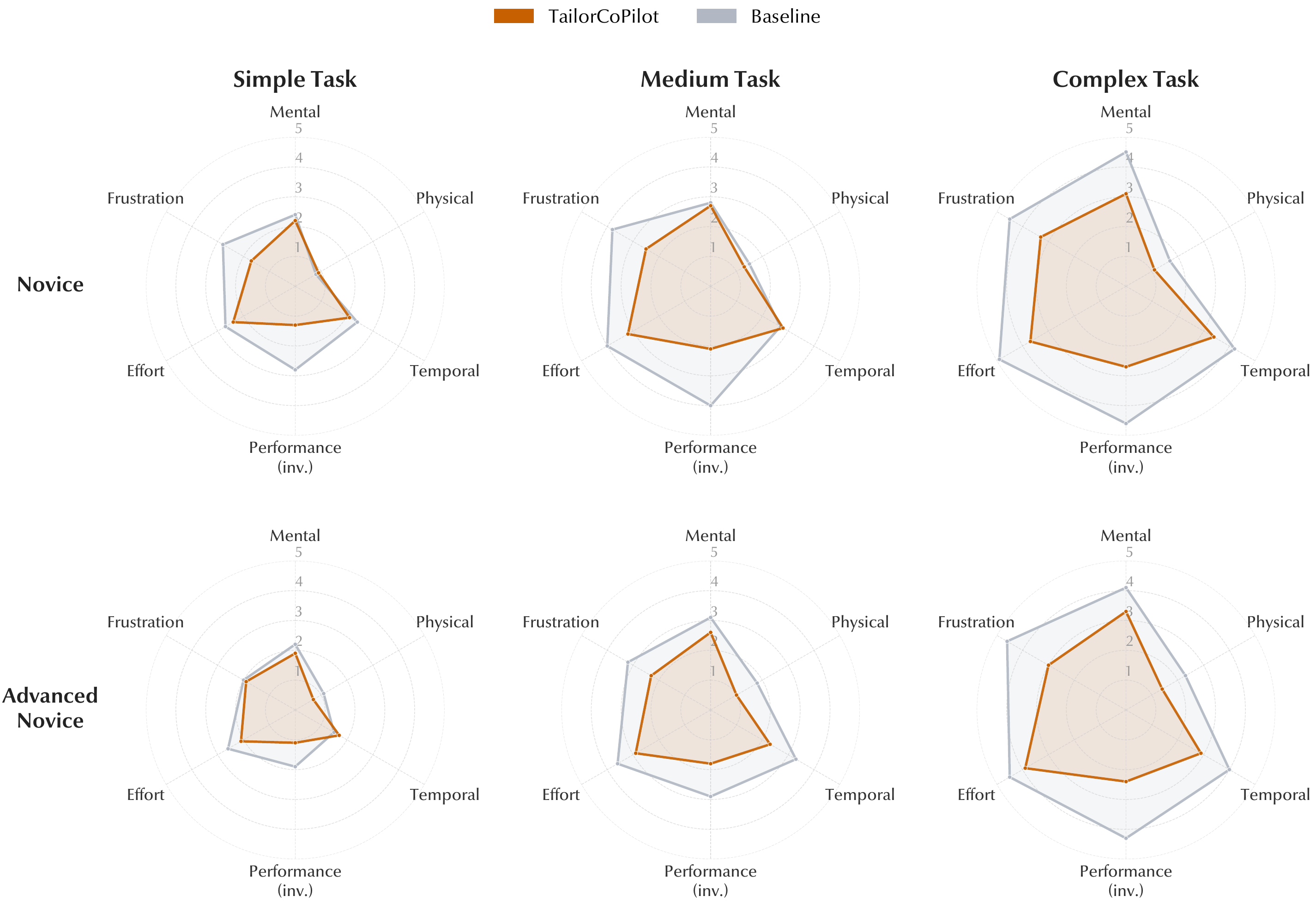}
\caption{Descriptive NASA-TLX profiles by workflow, task difficulty, and expertise group. Radar plots show mean ratings on the six NASA-TLX dimensions for novice and advanced novice participants under TailorCoPilot and the baseline. Across both expertise groups, the profiles suggest generally lower perceived workload under TailorCoPilot.}
\label{fig:app-nasa}
\Description{Six radar plots comparing NASA-TLX ratings for TailorCoPilot and the baseline across novice and advanced novice participants and across simple, medium, and complex tasks. Each plot shows the six NASA-TLX dimensions of mental demand, physical demand, temporal demand, inverted performance, effort, and frustration. Across most panels, the TailorCoPilot profile is smaller than the baseline profile, indicating lower perceived workload overall.}
\end{figure*}

\re{This section reports additional model statistics and interaction tests.

For task success, logistic regression showed that none of the tested interactions involving Workflow reached significance (all $p > .20$).

For completion time, the main effect of Workflow was significant, $F(1, 238) = 14.72$, $p < .001$, indicating a 41.7\% reduction in completion time with TailorCoPilot compared to the baseline workflow. The Workflow $\times$ Difficulty interaction was also significant, $F(2, 238) = 4.11$, $p = .018$. All other interactions involving Workflow were not significant (all $p > .30$).

For artifact quality, the main effect of Workflow was significant, $F(1, 238) = 11.94$, $p < .001$, with an estimated marginal mean difference of $1.85$ (95\% CI $[0.93, 2.77]$) on the artifact quality scale in favor of TailorCoPilot. No significant interactions were found (all $p > .24$).

For NASA-TLX, the main effect of Workflow was significant, $F(1, 238) = 30.25$, $p < .001$, indicating lower perceived workload with TailorCoPilot, corresponding to a Cohen's $d$ of 0.71. The Workflow $\times$ Expertise interaction was not significant ($p > .30$). Figure~\ref{fig:app-nasa} presents descriptive NASA-TLX profiles by expertise group and difficulty level using an adapted 5-point scale. Across expertise levels and difficulty conditions, TailorCoPilot generally resulted in lower mean scores across all six NASA-TLX dimensions compared with the baseline.}

\subsection{Additional Cases}
\begin{figure*}[t]
\centering
\includegraphics[width=0.695\linewidth]{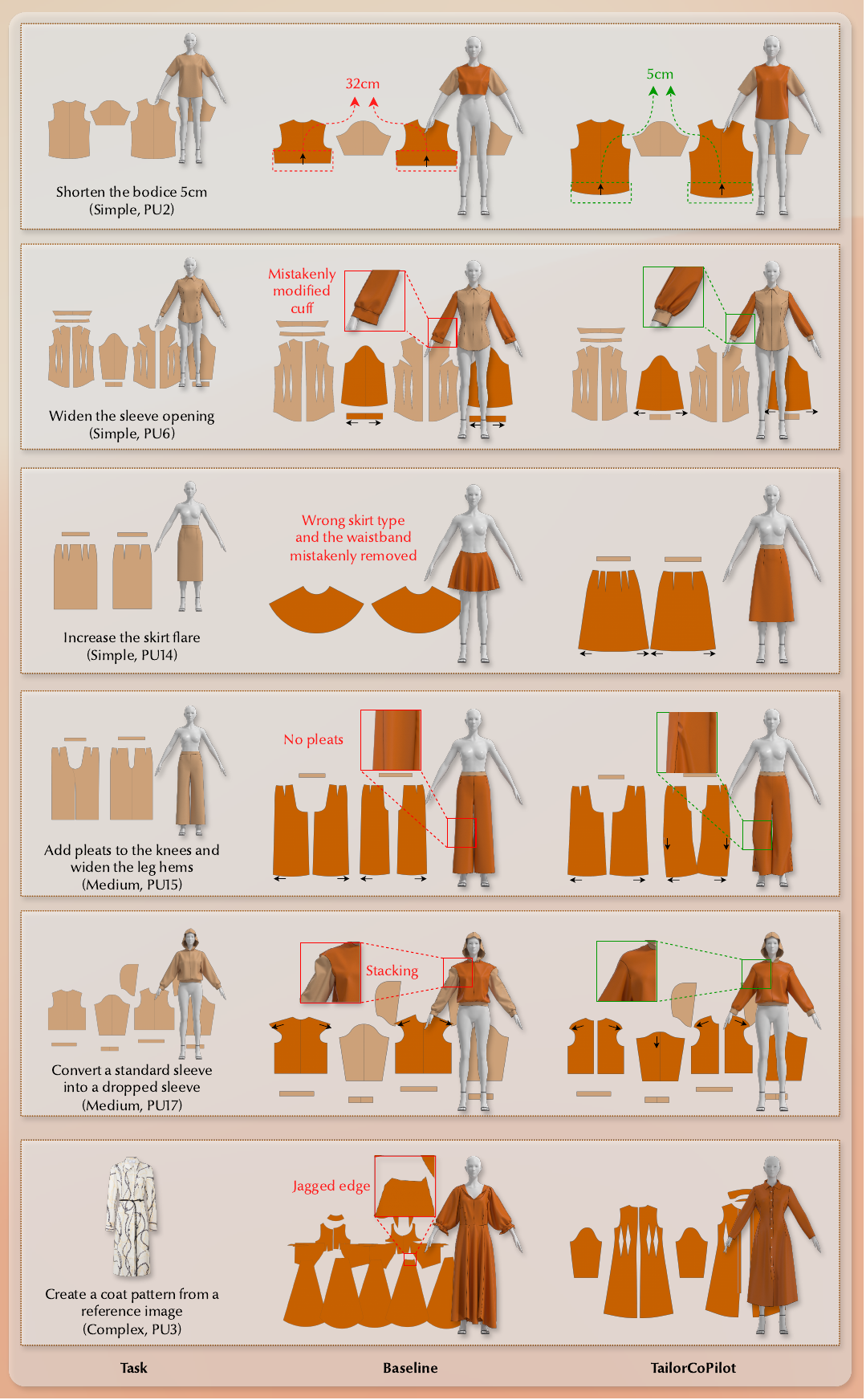}
\caption{Qualitative comparison across simple, medium, and complex tasks. Each row presents the input (base patterns or reference images), baseline outputs with typical errors, and improved results from TailorCoPilot.}
\label{fig:app-result}
\Description{
Six-row qualitative comparison of baseline and TailorCoPilot results across simple, medium, and complex tasks. Each row shows the task input on the left, a baseline output with a typical failure in the middle, and an improved TailorCoPilot result on the right. The examples highlight common baseline errors such as incorrect change magnitude, editing the wrong garment part, producing the wrong garment type, missing pleats, stacking artifacts, and jagged edges, while TailorCoPilot produces outputs that better match the intended edits and reference design.
}
\end{figure*}

We include additional qualitative examples selected as representative cases from successful trials to illustrate typical system behavior across different task complexities, as shown in Figure~\ref{fig:app-result}.

\bibliographystyle{ACM-Reference-Format}
\bibliography{sample-base}